\documentclass[prd,aps,twocolumn,preprintnumbers, showpacs, nofootinbib,superscriptaddress,notitlepage]{revtex4-1}
\usepackage{mathrsfs}
\usepackage{amsfonts}
\usepackage{amsmath}
\usepackage{slashed}
\usepackage{array}
\usepackage{verbatim}
\usepackage{epsfig}
\usepackage{graphicx}
\usepackage{color}
\usepackage[dvipsnames]{xcolor}

\newcommand{\beq}{\begin{eqnarray}}
\newcommand{\eeq}{\end{eqnarray}}

\def\be{\begin{equation}}
\def\ee{\end{equation}}
\def\bea{\begin{eqnarray}}
\def\eea{\end{eqnarray}}

\begin{document}

\title{RI/MOM renormalization of the quasi-PDF in lattice regularization}

\author{\includegraphics[scale=0.15]{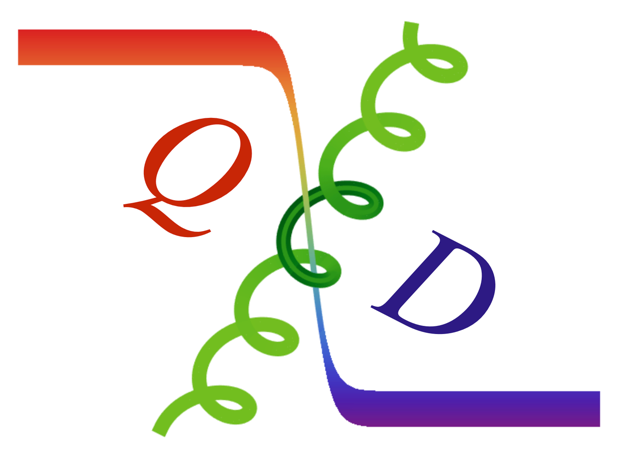}\\
Kuan Zhang}
\affiliation{University of Chinese Academy of Sciences, School of Physical Sciences, Beijing 100049, China}
\affiliation{CAS Key Laboratory of Theoretical Physics, Institute of Theoretical Physics, Chinese Academy of Sciences, Beijing 100190, China}

\author{Yuan-Yuan Li}
\affiliation{Nanjing Normal University, Nanjing, Jiangsu, 210023, China}

\author{Yi-Kai Huo}
\affiliation{Physics Department, Columbia University, New York, NY 10027, USA}

\author{Andreas Sch\"afer}
\affiliation{Institut f\"{u}r Theoretische Physik, Universit\"{a}t Regensburg, D-93040 Regensburg, Germany}

\author{Peng Sun}
\email{06260@njnu.edu.cn}
\affiliation{Nanjing Normal University, Nanjing, Jiangsu, 210023, China}

\author{Yi-Bo Yang}
\email{ybyang@itp.ac.cn}
\affiliation{CAS Key Laboratory of Theoretical Physics, Institute of Theoretical Physics, Chinese Academy of Sciences, Beijing 100190, China}
\affiliation{School of Fundamental Physics and Mathematical Sciences, Hangzhou Institute for Advanced Study, UCAS, Hangzhou 310024, China}
\affiliation{International Centre for Theoretical Physics Asia-Pacific, Beijing/Hangzhou, China}

\collaboration{\bf{$\chi$QCD Collaboration}}

\date{\today}

\begin{abstract}
  We analyze the lattice spacing dependence for the pion unpolarized matrix element of a quark bilinear operator with Wilson link (quasi-PDF operator) in the rest frame, using 13 lattice spacings ranging from 0.032 fm to 0.121 fm. We compare results for three different fermion actions with or without good chiral symmetry on dynamical gauge ensembles from three collaborations. This investigation is motivated by the fact that the gauge link generates an $1/a$ divergence, the cancelation of which in many ratios can be numerically tricky. Indeed, our results show that this cancelation deteriorates with decreasing lattice spacing, and that the RI/MOM method leaves a linearly divergent residue for quasi-PDFs. We also show that in the Landau gauge the interaction between the Wilson link and the external state results in a linear divergence which depends on the discretized fermion action.
\end{abstract}

\maketitle

\section{Introduction}


Parton Distribution Functions (PDFs) play a key role for most processes in high energy and hadron physics, ranging, e.g., from the search for New Physics to the detailed analysis of nucleon properties.
Therefore, it is one of the most important tasks of lattice QCD to determine PDFs by first principle theory calculations. While the calculation of PDF  Mellin moments, which can be expressed as matrix elements of local operators has a long, high-profile history, the calculation of their full $x$-dependence, which requires the evaluation of non-local correlators became only feasible in recent years. Among the different suggested approaches, the quasi-PDF method based on Large Momentum Effective Theory (LaMET)~\cite{Ji:2013dva} has been widely used, because it is relatively direct and has no model dependence and proved to be quite successful. For example, iso-vector unpolarized PDFs have been calculated using different types of gauge and fermion actions down to the physical pion mass~\cite{Alexandrou:2018pbm,Lin:2018qky,Hua:2020gnw}. A crucial topic which was not yet studied in great detail is the continuum extrapolation of quasi-PDFs. Controlling the continuum limit is already usually the main challenge for practicle lattice calculations of Mellin moments of PDFs. Because quasi-PDFs are far more delicate objects one has to expect that for them this task will be even more demanding, especially in combination with RI/MOM renormalization. At the same time, taking the continuum limit is indispensable for obtaining physical results, and therefore, intensifying such studies is the next logical  step in the further development of the LaMET approach. 

LaMET starts from the quasi-PDF operator $O_{\Gamma}(z)=\bar{\psi}(0) \Gamma U(0,z) \psi (z)$ with spatial gauge link $U(0, z) = \exp(-ig\int_0^{z} dz' A_z(z'))$, and connects its renormalized nucleon matrix element to PDFs through factorization theorems~\cite{Ji:2013dva,Ji:2014gla,Ma:2014jla,Ma:2017pxb,Izubuchi:2018srq}. However, the bare $O_{\Gamma}(z)$ in lattice regularizations contains an $1/a$ term
\bea\label{eq:1-loop}
O_{\Gamma}(z)=\Gamma\big(1+g^2(\gamma \textrm{log}(p^2a^2)-m_{-1}\frac{z}{a})+......),
\eea
at 1-loop level~\cite{Alexandrou:2017huk} with lattice spacing $a$, and can deviate from its tree level value exponentially at either large $z$ or small $a$ when higher order effects are summed over. Thus, non-perturbative renormalization with accurate cancellation on the linear divergence is essential for the quasi-PDF, to ensure the existence of a finite continuum limit.

Studies in the continuum~\cite{Ji:2015jwa, Ji:2017oey,Ishikawa:2017faj,Green:2017xeu} suggest that the quasi-PDF operator is multiplicatively renormalizable and that the linear divergence just comes from the Wilson link and is independent of the external quark or hadron state. Therefore, RI/MOM renormalization~\cite{Martinelli:1994ty} is believed to be a good candidate to remove the linear divergence~\cite{Constantinou:2017sej,Green:2017xeu,Chen:2017mzz,Stewart:2017tvs,Alexandrou:2017huk,Alexandrou:2020qtt,Lin:2020fsj}.
{However, an accurate non-perturbative examination of this cancellation in lattice regularization is still missing.}

To close this gap, we study the RI/MOM renormalized pion quasi-PDF matrix element in the rest frame, for 13  lattice spacings ranging from from 0.032 fm to 0.121 fm using different quark and gluon actions. Our results show that the cancelation deteriorates with decreasing lattice spacing, and that the RI/MOM method leaves a linearly divergent residue for quasi-PDFs. We also show that in the Landau gauge the interaction between the Wilson link and the external state results in a linear divergence which depends on the discretized fermion action.

\section{Numerical setup}

When we consider the quasi-PDF nucleon matrix element in the moving frame, the gap $\delta m$ between the ground and first excited state decreases with increasing momentum, requireing a large source/sink separation $t_{\textrm{sep}}$ to eliminate excited state contaminations. At the same time, the signal to noise ratio also decreases exponentially with $t_{\textrm{sep}}$, reducing our ability to identify the lattice spacing dependence from data.

Using the pion rest frame can, therefore, be the better choice as it avoids this problem. It is known that the first excited state of the pion lies higher than 1GeV, and that the signal to noise ratio is almost independent of $t_{\textrm{sep}}$. Thus we can choose $t_{\textrm{sep}}$ to be half of the temporal lattice length $T$ ($T$ is mostly larger than 6 fm for the ensembles we use), and then obtain the ground state matrix element  $h_{\pi}=\langle \pi| O_{\gamma_t}(z)|\pi \rangle$ with high accuracy:
\begin{align}\label{eq:hadron}
&R_{\pi}(T/2,t,z;a)\nonumber\\
&\equiv\frac{\langle O_\pi(T/2)\sum_{\vec{x}}\big(O_{\Gamma}(z;(\vec{x},t))+O_{\Gamma}(z;(\vec{x},T-t))\big)O_{\pi}^{\dagger}(0)\rangle}{\langle O_{\pi}(T/2)O_{\pi}^{\dagger}(0)\rangle}\nonumber\\
&=h_{\pi,\Gamma}(z)+{\cal O}(e^{-\delta m t})+{\cal O}(e^{-\delta m (T/2-t)})+{\cal O}(e^{-\delta m T/2}),
\end{align}
where $O_{\pi}$ is the pion interpolating field. Note that the denominator includes both forward and backward two point functions. These are  needed
because of the possibility to loop around the lattice in the temporal direction for periodic boundary conditions.

In RI/MOM renormalization, we define the renormalization constant as~\cite{Liu:2018uuj},
\begin{align}\label{eq:Z_ri}
Z_{\gamma_t}(z,\mu)=\frac{Z_q(\mu)}{\textrm{Tr}[\gamma_t\langle q(p)|O_{\gamma_t}(z)|q(p)\rangle]_{p^2=-\mu^2, p_z=p_t=0}},
\end{align}
where $|q(p)\rangle$ is the off-shell quark state with external momentum $p$, and $Z_q$ is defined from the pion matrix element of a local vector current,
\begin{align}\label{eq:Z_q}
Z_q(\mu)=\frac{\textrm{Tr}[\gamma_t\langle q(p)|O_{\gamma_t}(0)|q(p)\rangle]_{p^2=-\mu^2, p_z=p_t=0}}{h_{\pi,\gamma_t}(0)}.
\end{align}
 When the Landau gauge fixed volume ($V$) source~\cite{Chen:2017mzz} is used in the calculation, the statistical uncertainty is suppressed by a factor $1/\sqrt{V}$ compared to the point source case. Note that the $Z_q$ definition used here is exactly the same as $\tilde{Z}_q\equiv \textrm{Tr}[p\!\!\!/S^{-1}]/p^2$ from the quark propagator in dimensional regularization~\cite{Gracey:2003yr}, but the discretization error is much smaller than for $\tilde{Z}_q$ in lattice regularization as shown in Ref.~\cite{Chang:2021vvx}.

We can apply $Z_{\gamma_t}$ to the bare pion matrix element $h_{\pi,\gamma_t}(z)$ to obtain the non-perturbatively renormalized and normalized matrix element at a given RI/MOM scale $\mu$,
\begin{align}\label{eq:h^r}
h^{r}_{\pi,\gamma_t}(z,\mu)=Z_{\gamma_t}(z,\mu)h_{\pi,\gamma_t}(z).
\end{align}

In order to check if the linear divergence is related to the fermion action, we compare results for two kinds of fermion actions in this work: the clover (CL) and overlap (OV) actions.  The clover action is computationally relatively cheap and widely used in quasi-pdf calculations, while the overlap action is much more expensive but conserves chiral symmetry.

The clover fermion action is
\begin{align}\label{eq:clover}
&S^{\textrm{w}}_{q}=\sum_{x,y}\bar{\psi}(x)D_\textrm{w}(m^{\textrm{w}}_q;x,y)\psi(y),\nonumber\\
&D_\textrm{w}(m^{\textrm{w}}_q;x,y)=\frac{1-\gamma_{\mu}}{2a}U(x,x+\hat{n}_{\mu})\delta_{x+\hat{n}_{\mu},y}\nonumber\\
&+\frac{1+\gamma_{\mu}}{2a}U(x,x-\hat{n}_{\mu})\delta_{x-\hat{n}_{\mu},y}-(\frac{4}{a}+m^{\textrm{w}}_q))\delta_{x,y},
\end{align}
with an additional ``clover'' term,
\begin{align}\label{eq:clovr}
&S^{\textrm{cl}}_{q}=S^{\textrm{w}}_{q}+ac_{\textrm{sw}}\sum_{x}\bar{\psi}(x)\sigma_{\mu\nu} F_{\mu\nu}(x)\delta_{x,y}\psi(y),
\end{align}
where $\hat{n}_{\mu}$ is the unit vector along the $\mu$ direction, $m^\textrm{w}_q-m^{\textrm{cri}}$ is the multiplicatively renormalizable bare quark mass, and $m^{\textrm{cri}}$ is the value of the bare quark mass for which the pion mass vanishes. $m^{\textrm{cri}}$ is ${\cal O}(\frac{\alpha_s}{a})$ at leading order for the Wilson action and always negative. It can be reducsed to ${\cal O}(\frac{\alpha^2_s}{a})$ (but usually still negative) with a clover coefficient $c_{\textrm{sw}}=1+{\cal O}(\alpha_s)$. It can be further suppressed by applying gauge link smearing and/or fine-tuning of $c_{\textrm{sw}}$. 

To eliminate $m^{\textrm{cri}}$ exactly one can use chiral fermions which satisfies the Ginsburg-Wilson relation $D_{\textrm{ov}}\gamma_5+\gamma_5D_{\textrm{ov}}=\frac{a}{\rho}D_{\textrm{ov}}\gamma_5D_{\textrm{ov}}$~\cite{Ginsparg:1981bj}. For example, Refs.~\cite{Narayanan:1994gw,Neuberger:1997fp,Chiu:1998eu,Liu:2002qu} define overlap fermion by
\begin{align}
S^{\textrm{ov}}_{q}&=\sum_{x,y}\bar{\psi}(x) D_{\textrm{ov}}(x,y)\psi(y),\\
 D_{\textrm{ov}}&=\rho\Big(1+\frac{D_\textrm{w}(-\rho)}{\sqrt{D^{\dagger}_\textrm{w}(-\rho)D_\textrm{w}(-\rho)}}\big)\Big),\nonumber
\end{align}
where $-\rho$ should be smaller than $m^{\textrm{cri}}$, to make $D_{\textrm{ov}}$ to be the same as the standard Dirac operator in the continuum limit. The chiral fermion propagator is defined by $D_{\textrm{ov}}$,
\begin{align}
\!\!\!\frac{1}{D_c+m^{\textrm{ov}}_q}=\frac{1}{\frac{D_{\textrm{ov}}}{1-\frac{1}{2\rho}D_{\textrm{ov}}}+m^{\textrm{ov}}_q}=\frac{1-\frac{1}{2\rho}D_{\textrm{ov}}}{D_{\textrm{ov}}+m^{\textrm{ov}}_q(1-\frac{1}{2\rho}D_{\textrm{ov}})},
\end{align}
where $D_c$ satisfies the relation $D_c\gamma_5+\gamma_5D_c=0$.  Then, $m_q\rightarrow 0$ make the pion mass vanish without any fine-tuning.

\begin{table*}[htbp]
  \centering
  \begin{tabular}{c|ccrc|cc|cc|c||c|ccrc|c}
  \toprule
tag &  $6/g^2$ & $L$ & $T$ & $a(\mathrm{fm})$ & $c_{\mathrm{sw}}$ & $m^{\textrm{w}}_{q}a$ & $c'_{\mathrm{sw}}$ & $m^{\textrm{w}'}_{q}a$ & $m^{\textrm{ov}}_{q}a$& tag &  $6/g^2$ & $L$ & $T$ & $a(\mathrm{fm})$ & $m^{\textrm{ov}}_{q}a$\\
\hline
MILC12 &  3.60 & 24 & 64 & 0.1213(9) & 1.0509   & -0.0695 & 1.31 & 0.010 & 0.015 & RBC11 & 2.13 & 24 & 64 & 0.1105(3)& 0.015\\
\hline
MILC09 &  3.78 & 32 & 96 & 0.0882(7) & 1.0424  & -0.0514 & -- & --  & 0.011 & RBC08 & 2.25 & 32 & 64 & 0.0828(3)  & 0.011\\
\hline
MILC06 &  4.03 & 48 & 144 & 0.0574(5) & 1.0349  & -0.0398 & 1.25 & 0.0014 & 0.008 & RBC06 & 2.37 & 48 & 96 &0.0627(3) & 0.008\\
\hline
MILC04 &  4.20 & 64 & 192 & 0.0425(4) & 1.0314  & -0.0365 & -- & -- & 0.005\\
\hline
MILC03 &  4.37 & 96 & 288 & 0.0318(3) & 1.0287  & -0.0333 & 1.26 & 0.0030 & 0.0035 \\
  \end{tabular}
      \begin{tabular}{c|ccrc|cc||c|ccrc|cc}
  \toprule
tag &  $6/g^2$ & $L$ & $T$ & $a(\mathrm{fm})$ & $c_{\mathrm{sw}}$ & $m^{\textrm{w}}_{q}a$ & tag &  $6/g^2$ & $L$ & $T$ & $a(\mathrm{fm})$ & $c_{\mathrm{sw}}$ & $m^{\textrm{w}}_{q}a$\\
\hline
CLS10 &  3.34 & 24 & 48 & 0.0980(12) & 2.06686   & -0.3437  &CLS08 &  3.40 & 32 & 96 & 0.0854(10) &  1.98625  & -0.3468 \\
\hline
CLS06 &  3.55 & 48 & 128 & 0.0644(08) &  1.82487 & -0.3525 &CLS05 &  3.70 & 48 &128 & 0.0500(07) &  1.70477  & -0.3521 \\
\hline
CLS04 &  3.85 & 64 &192 & 0.0390(06) &  1.61281  & -0.3478 \\
\hline
  \end{tabular}
  \caption{Setup of the ensembles, including the bare coupling constant $g$, lattice size $L^3\times T$ and lattice spacing $a$. $m^\textrm{w}_q$ and $m^{\textrm{w}'}_q$ are the bare quark masses using the clover fermion action with the two clover coefficient  $c_{\mathrm{sw}}$ and $c'_{\mathrm{sw}}$ respectively, and $m^{\textrm{ov}}_q$ is the bare quark mass of the overlap fermion. The pion masses in all cases are in the range 310-360 MeV.}
  \label{tab:lattice}
\end{table*}

\section{Results}

We use three sets of ensembles:

(i) $N_f=2+1+1$ highly improved staggered fermions (HISQ) and one-loop Symanzik improved gauge fields from the MILC Collaboration~\cite{Bazavov:2012xda} at five lattice spacings (MILC12--MILC03).

(ii) $N_f=2+1$ domain wall fermions (DWF) and Iwasaki gauge fields from the RBC/UKQCD collaboration~\cite{Blum:2014tka} at three lattice spacings (RBC11--RBC06).

(iii) $N_f=2+1$ clover fermions and Luescher-Weisz (equivalent to Symanzik) gauge fields from the CLS collaboration~\cite{Bruno:2014jqa} at five lattice spacings (CLS10--CLS04).

For the MILC/RBC gauge configurations we apply 1-step HYP smearing~\cite{Hasenfratz:2001hp}. For the overlap fermions we use $\rho=1.5/a$ and we use two clover coefficients: one is the tadpole improved tree level coefficient $c_{\mathrm{sw}}$ which is very close to one after the configuration is HYP smeared (CL), and the other is $c'_{\mathrm{sw}}\sim 1.3$ which gives a critical quark mass $m^{\textrm{cri}}_{q}$ around zero (tuned-CL). The bare overlap quark mass $m_q^{\textrm{ov}}a$ is roughly proportional to the lattice spacing if we require the pion mass to be around 310 MeV, but $m^{\textrm{w}}_{q}a$ 
is always negative with $c_{\mathrm{sw}}\sim1$ unless we enlarge $c_{\mathrm{sw}}$ to $\sim$ 1.3. For the CLS configuration, we simply use the unitary setup (without any smearing of the action) in the clover fermion calculations. Since most of the CLS ensembles use the open boundary condition except CLS10, we just use the 1/3 of the time slices in the middle of the lattice to avoid lattice artifacts for the volume source used for the RI/MOM renormalization constants. 

\begin{figure}[tbph]
  \centering
  \includegraphics[width=8cm]{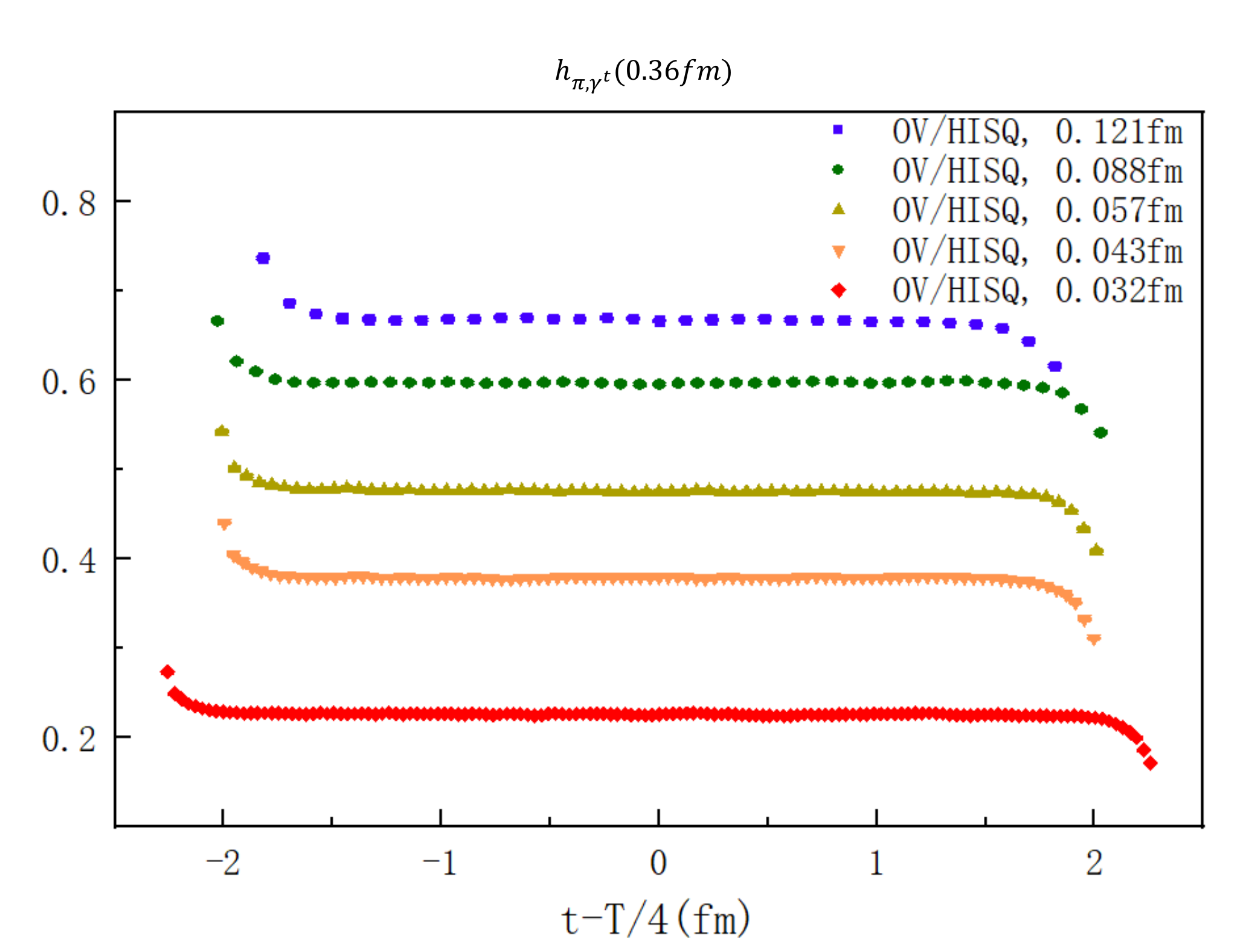}
  \caption{The bare ratios $R_{\pi}(T/2,t,z;a)$ defined in Eq.~(\ref{eq:hadron}) for the pion using overlap (OV) fermions on HISQ sea ensembles at different lattice spacings. There are clear plateaus in the region $t-T/4\in$[-1,1] fm. }
  \label{fig:ratio0}
  \end{figure}

\begin{figure}[tbph]
  \centering
  \includegraphics[width=8cm]{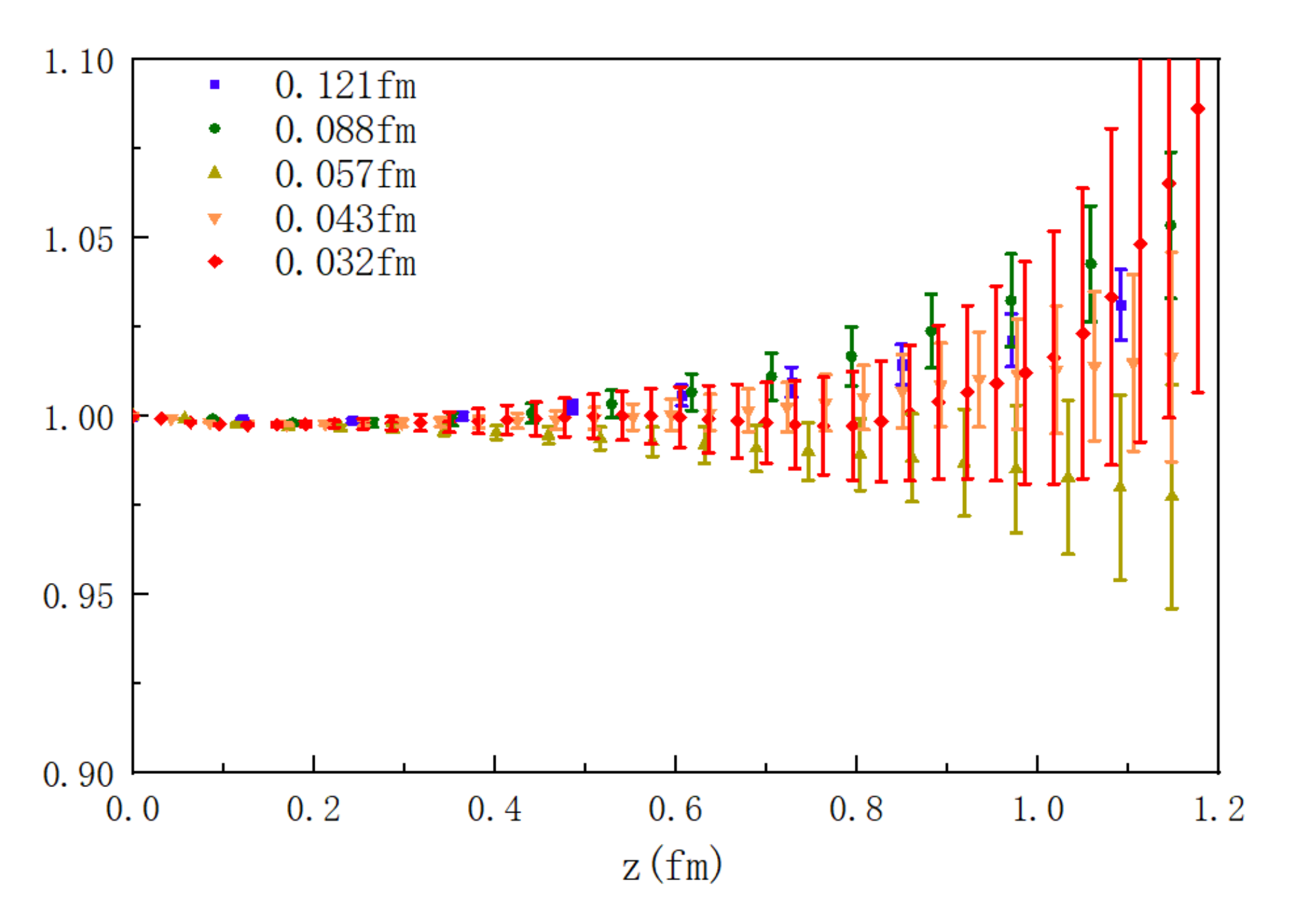}
  \caption{The ratios between the pion matrix element $h_{\pi,\gamma_t}(z)$ using clover (CL) fermions and overlap (OV) fermions, on the HISQ sea ensembles at different lattice spacings. The ratios are consistent with 1 within 2 $\sigma$ at all lattice spacings.}
  \label{fig:ratio1}
  \end{figure}

We start with the pion matrix element and the MILC ensemble using the overlap action, to demonstrate the elimination of excited state contaminations. As shown in Fig.~\ref{fig:ratio0}, the bare ratio $R_{\pi}(T/2,t,z\simeq 0.36 $~fm$;~a)$ defined in Eq.~(\ref{eq:hadron}) is independent of $t$ in the region $t-T/4\in$[-1,1] fm. Thus we average $R_{\pi}(T/2,t,z;a)$ in the region $t\in[T/8,3T/8]$ to get a precise estimate of the ground state matrix elements. We can also see that the ratio becomes exponentially smaller at smaller lattice spacings. Non-perturbative renormalization is essential to recover a reasonable continuum limit. A similar figure for the clover case can be found in our related work~\cite{Huo:2021rpe} which uses the same data sets. It is worths emphasizing that if we take the ratio between the pion matrix elements $h_{\pi,\gamma_t}(z)$ obtained using the clover fermion and overlap fermion, then the ratio is consistent with 1 within 2 $\sigma$, or in other words with at most 3\% difference for all $z<$ 1.0 fm as shown in Fig.~\ref{fig:ratio1}. This agrees with the previous impression that the linear divergence is independent of the discretized fermion action.

  \begin{figure}[tbp]
  \centering
\includegraphics[width=8cm]{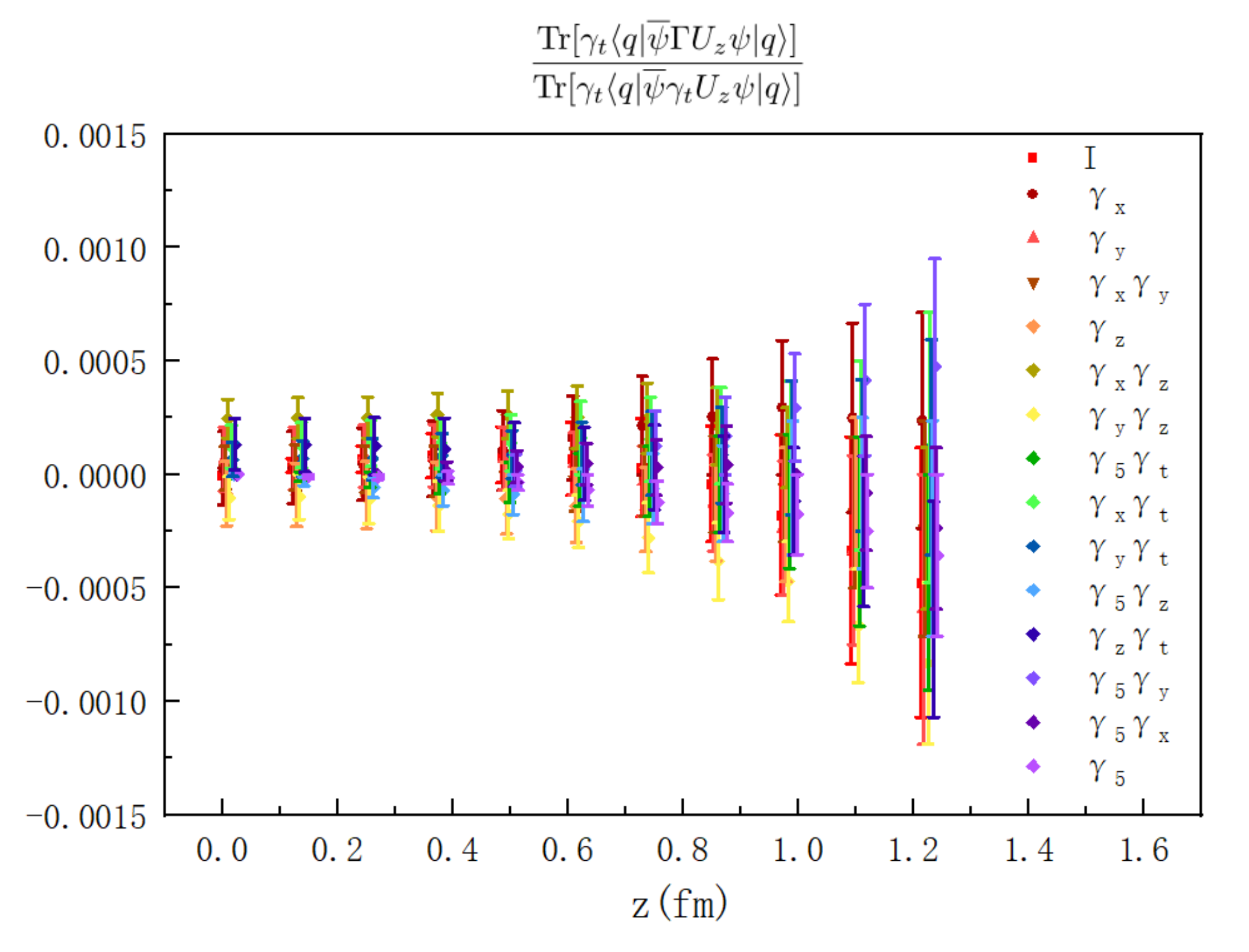}
\includegraphics[width=8cm]{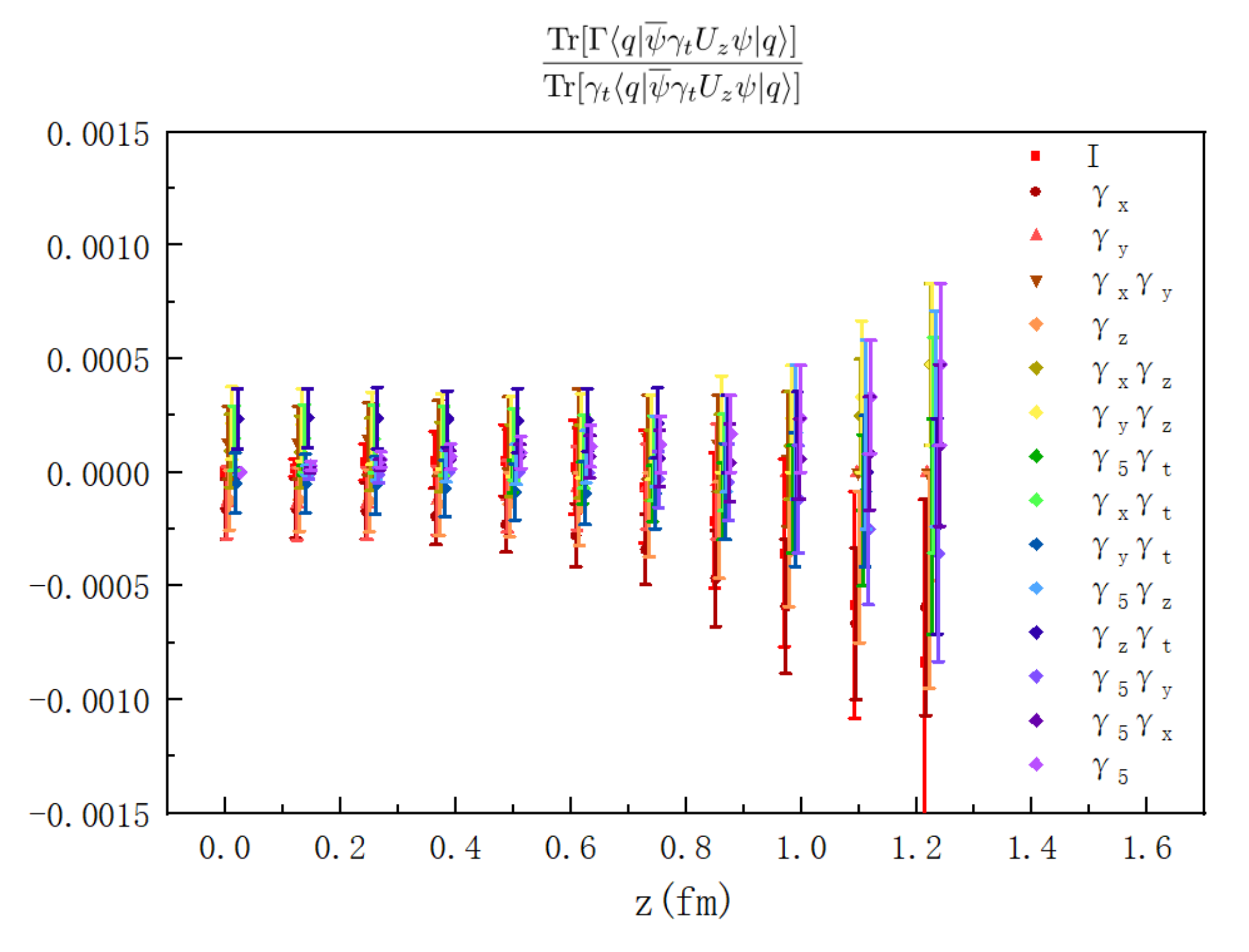}
    \caption{The real part of the relative mixing between $O_{\gamma_t}(z)$ and $O_{\Gamma\neq\gamma_t}(z)$, using clover fermions on the HISQ sea ensemble with $a$=0.121 fm. The upper panel shows the $O_{\gamma_t}$ matrix element with a $\Gamma\neq\gamma_t$ projector, and the lower panel shows the $O_{\Gamma\neq\gamma_t}$ matrix element with a $\gamma_t$ projector. In all cases the results are consistent with zero within less than 0.1\% statistical uncertainty.}
  \label{fig:mixing}
  \end{figure}

Based on a symmetry analysis and 1-loop calculation, we can decompose the amputated Green's function $\Lambda
_{\gamma_t}(z,p)\equiv \langle q(p)|O_{\gamma_t}(z)|q(p)\rangle$ into the following Lorentz structures:
\begin{align}\label{eq:mixing_pattern}
\Lambda_{\gamma_t}(z,p)=\tilde{F}_t(z,p)\gamma_t+\tilde{F}_z(z,p)\{\gamma_z z\}p_t+\tilde{F}_p(z,p)\frac{p_tp\!\!\!/}{p^2}.
\end{align}
This is equivalent to what was proposed in Ref.~\cite{Liu:2018uuj} but the form Eq.~(\ref{eq:mixing_pattern}) is more natural to avoid $p_z$ to appear in a denominator. When $z=0$, the $\tilde{F}_z$ term vanishes and Eq.~(\ref{eq:mixing_pattern}) becomes a matrix element of the local operator $\bar{q}\gamma_{\mu}q$. Both, the $\tilde{F}_{z}$ and $\tilde{F}_{p}$ terms vanish when $p_t=0$. 

At the same time, setting $p_z=0$ leads to a vanishing imaginary part of $F_t$, as shown in the 1-loop calculations~\cite{Constantinou:2017sej,Liu:2018uuj}. Since $h_{\pi,\gamma_t}(z)$ is real in the rest frame, using the RI/MOM momentum with $p_z=p_t=0$ is the optimal choice to avoid complications from a complex renormalization constant as well as operator mixings. Based on the numerical calculation, we verified that the renormalization constant $Z_{\gamma_t} $ is real and the mixing between $O_{\gamma_t}(z)$ and $O_{\Gamma\neq\gamma_t}(z)$ is consistent with zero within very small uncertainties. Fig.~\ref{fig:mixing} shows the real part of the operator mixing coefficients using clover fermions on the HISQ sea ensemble with $a$=0.121 fm. They are consistent with zero within 0.1\% relative statistical uncertainties in all cases.

It is also popular to require all momentum components to be non-zero to suppress discretization errors, and use the p-slash projection as proposed in Ref.~\cite{Stewart:2017tvs} when operator mixing is unavoidable. Since the conclusion is still unchanged with the p-slash projection, we place the related discussion in the appendix for the interested readers.

  \begin{figure}[tbp]
  \centering
\includegraphics[width=8cm]{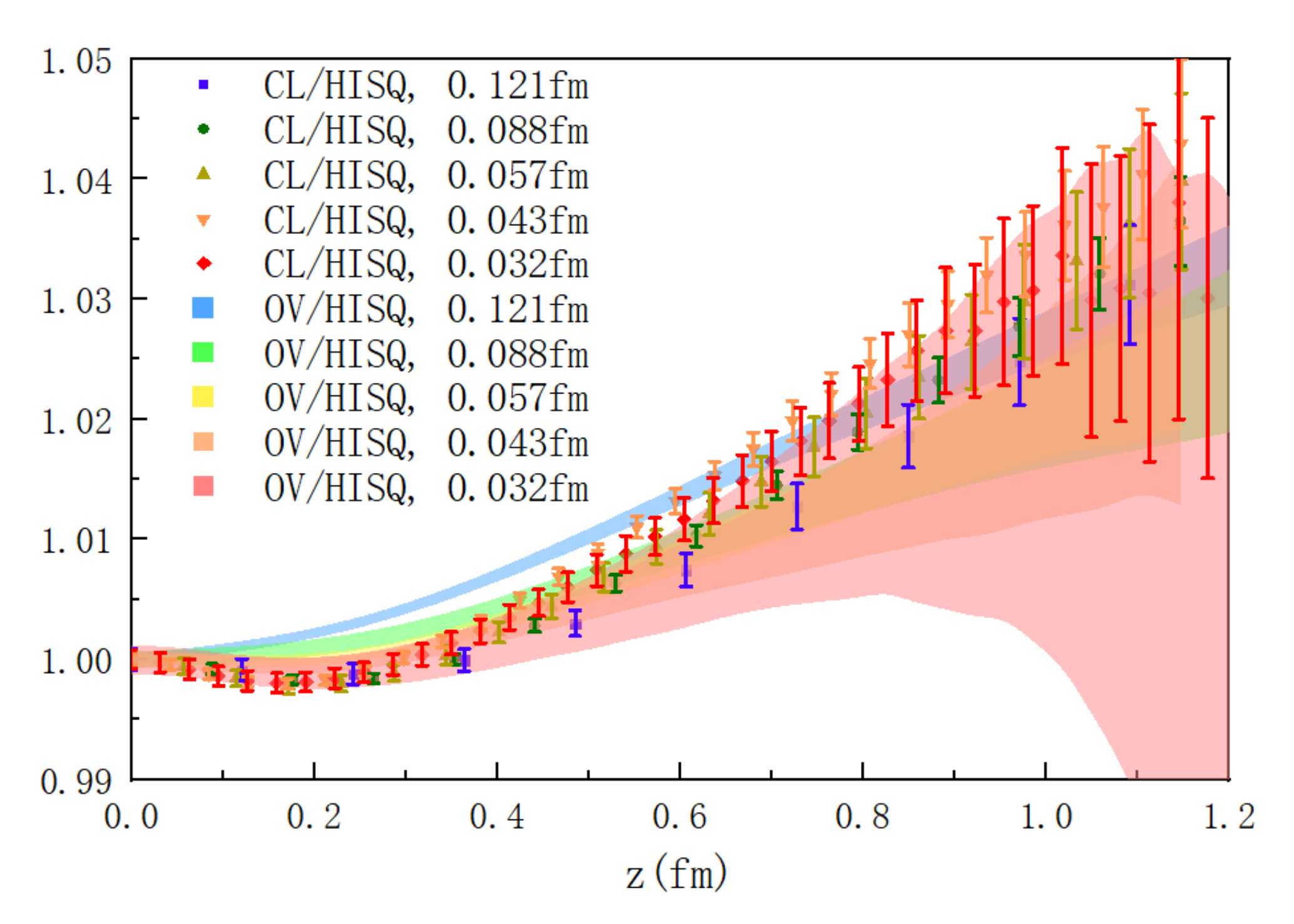}
\includegraphics[width=8cm]{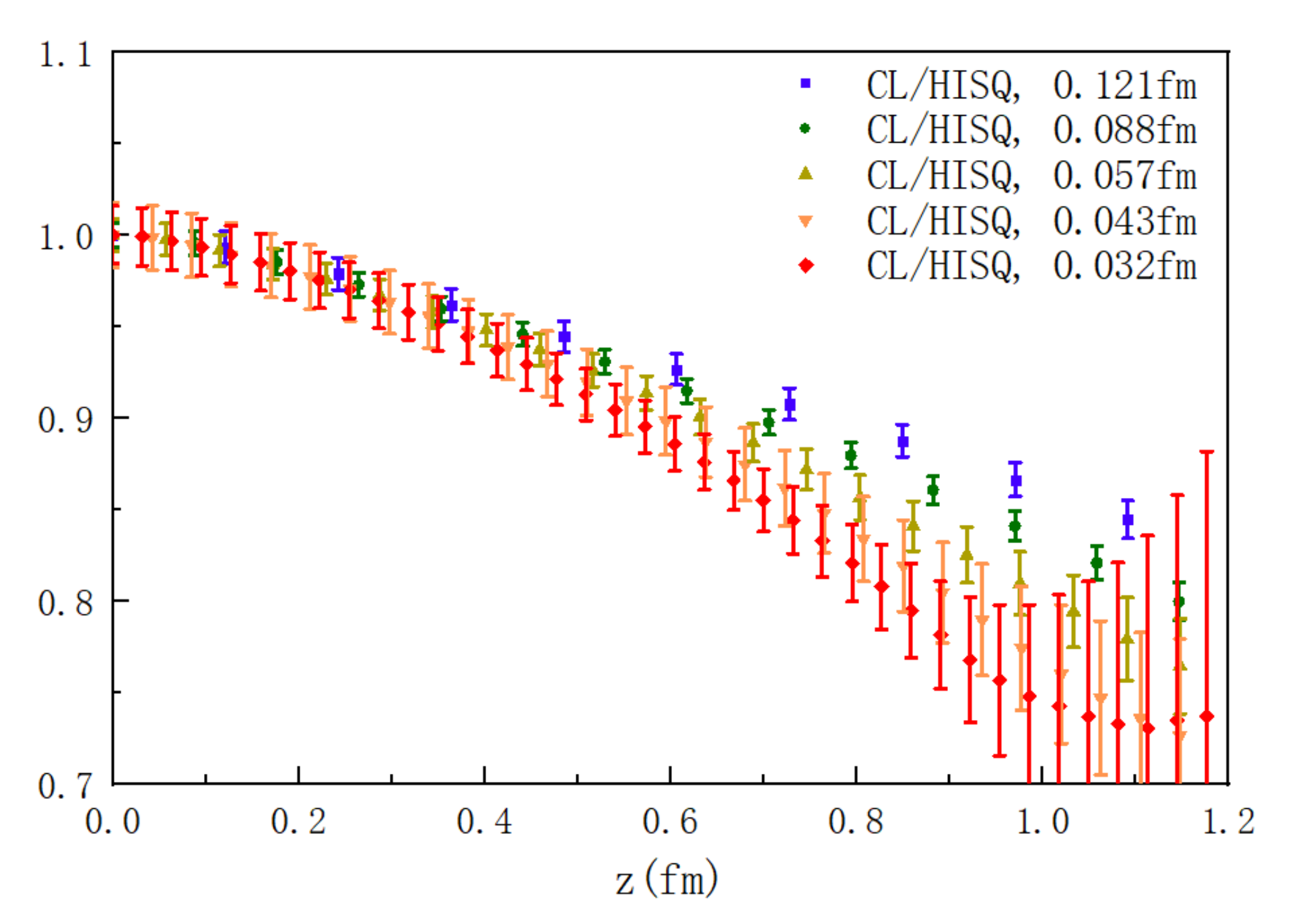}
    \caption{The $z$-dependent ratio $R_{\textrm{RI/MOM}}(z,\mu,\mu')$ defined in Eq.~(\ref{eq:ri_ratio}) with $\mu\simeq$ 3~GeV and  $\mu'\simeq$ 1.8~GeV (upper panel), and also $R_{\textrm{RI/MOM}}(z,\mu,\mu'')$ with $\mu''=$ 0~GeV (lower panel) using the given valence action and lattice spacing on the HISQ sea ensembles. The results with clover (CL) and overlap (OV) actions converge to the same continuum limit.}
  \label{fig:npr}
  \end{figure}

  As for the RI/MOM renormalization constant in this work, we use the momentum $p=2\pi(5,5,0,0)/L$ on the HISQ sea ensembles and tune $p_{x,y}$ on the DWF sea ensembles to make $\mu=\sqrt{p^2}$ to be the same on all ensembles within 6\%. As shown in the upper panel of Fig.~\ref{fig:npr} for the ratio of the renormalization constants,
\begin{align}\label{eq:ri_ratio}
R_{\textrm{RI/MOM}}(z,\mu,\mu')={Z_{\gamma_t}(z,\mu)}/{Z_{\gamma_t}(z,\mu')}
\end{align}
with $\mu'=\sqrt{-(p')^2}=1.8$ GeV ($p'=2\pi(3,3,0,0)/L$) on the HISQ sea ensembles using different valence fermion actions and lattice spacings,  the ratio just changes by about 2\% at $z$=1 fm regardless of the actions and lattice spacings used. Since the 6\% fluctuation of $\mu$ on different ensembles is much smaller than the difference between $\mu'$ and $\mu$, the systematic uncertainty due to this fluctuation will be much smaller than 2\% at $z$=1 fm and thus negligible. At the same time, Fig.~\ref{fig:npr} also shows that $R_{\rm RI/MOM}$ converges at small lattice spacing and is independent of the discretized fermion actions (clover or overlap).
 
However, the residual $z$ dependence indicates that the $\mu$ and $z$ dependences of $Z(z,\mu)$ are not independent. Thus it is natural to be curious about its behavior when $\mu$ decreases towards the on-shell limit. Thus we also consider the ratio with $\mu''=\sqrt{-(p'')^2}=0$ GeV using the clover valence fermion action on MILC ensembles. In the lower panel of  Fig.~\ref{fig:npr}, the $z$ dependence is 10 times stronger than for $\mu'\simeq1.8$ GeV, but the ratios at different lattice spacings are still consistent with each other. We conclude from the results in the two panels of Fig.~\ref{fig:npr} that the UV divergence of the RI/MOM renormalization constant is independent of the external off-shell momenta.

  \begin{figure}[tbp]
  \centering
  \includegraphics[width=8cm]{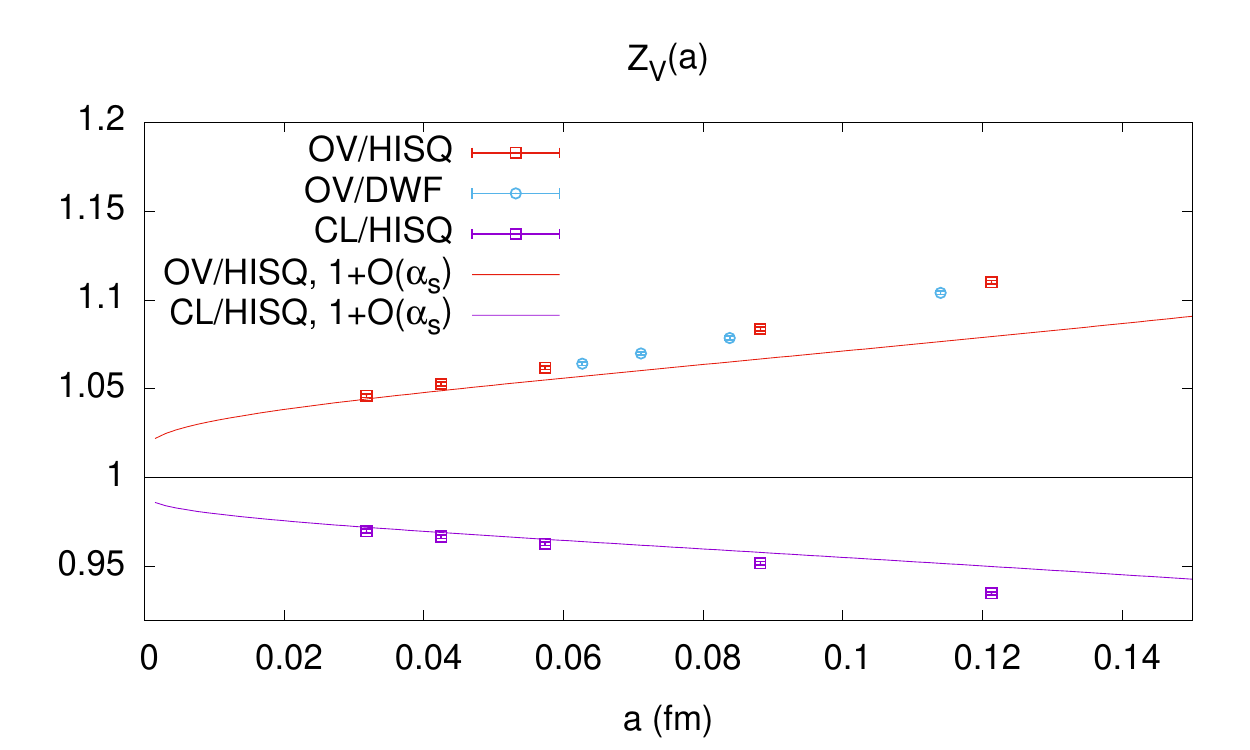}
  \caption{The normalization factor $Z_{\gamma_t}(0,\mu)$, which contains both $\alpha_s$ and also $a^2$  corrections as shown in Ref~\cite{Bazavov:2018omf}. The lines correspond to the renormalization factors with the ${\cal O}(a^2)$ correction subtracted.}
  \label{fig:zv}
  \end{figure}  

Another thing we need to mention here are the normalization conditions Eqs.~(\ref{eq:Z_ri}--\ref{eq:Z_q}) of the vector current. Based on the definition of the RI/MOM scheme, the quark self energy is defined from the charge conservation condition which requires the renormalized local vector charge to be unity, $h^r_{\pi, \gamma_t}=1$. In Fig.~\ref{fig:zv}, we show the normalization factors $Z_{\gamma_t}(0,\mu)$ we obtained with overlap fermions on HISQ sea ensembles (red boxes) and DWF sea ensembles (blue dots), and also for clover fermions on HISQ sea ensembles (purple boxes). After performing simple fits for the $\alpha_s$ and $a^2$ corrections~\cite{Bazavov:2018omf}, we can also show the lattice spacing dependence of the normalization factor with the ${\cal O}(a^2)$ correction subtracted, using the red line for the overlap case and purple line for the clover case. It is clear that the normalization we used  with the $\alpha_s$ correction approaches 1 smoothly and does not introduce any additional linear divergence.

  \begin{figure}[tbp]
  \centering
  \includegraphics[width=8cm]{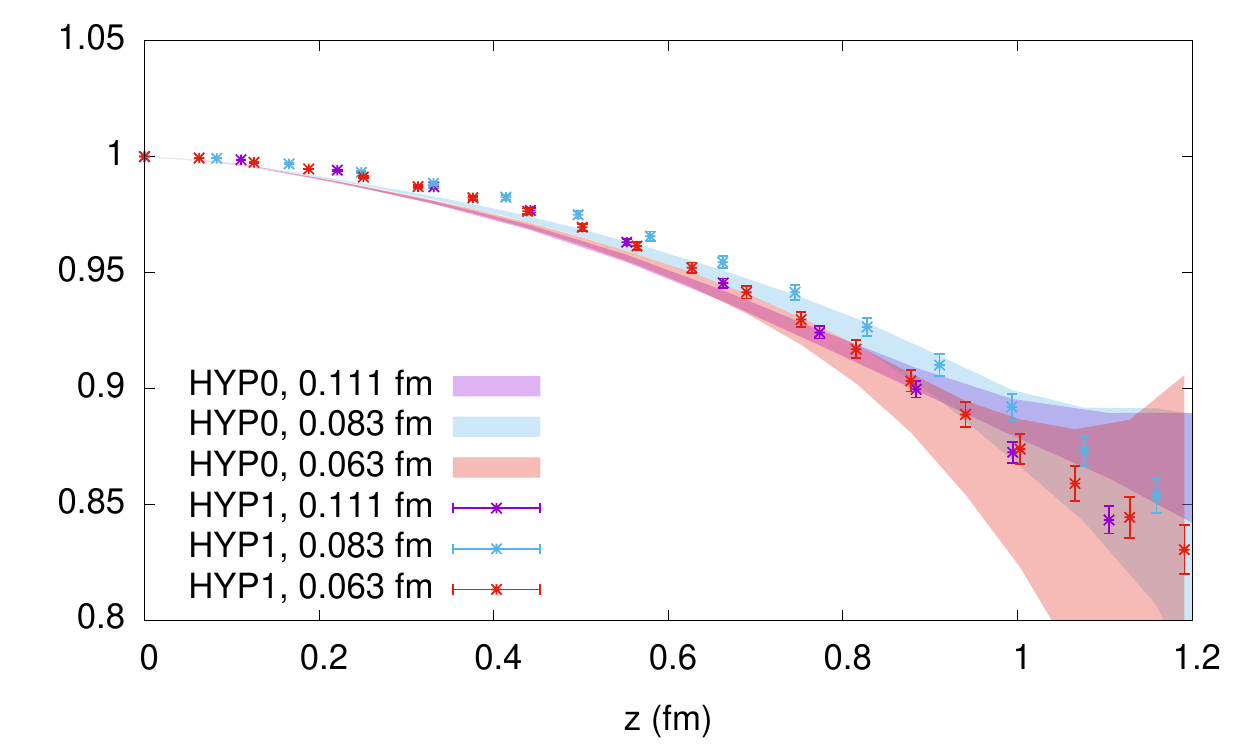}
  \caption{The RI/MOM renormalized pion matrix element of the vector current, using overlap (OV) fermions on DWF sea ensembles. The data points correspond to the cases with 1-step of the HYP smearing, and the bands are the results without HYP smearing of the Wilson link.}
  \label{fig:overlap1}
  \end{figure}

\subsection{RI/MOM renormalized pion matrix element with the overlap action}

We start from the overlap fermion case which is free of a possible linear divergence due to a possible mistuning of $m^{\rm cri}$.  Fig.~\ref{fig:overlap1} compares the RI/MOM renormalized pion matrix elements with the vector current $h^r_{\pi, \gamma_t}(z,a)$, using overlap fermion on the domain wall sea with and without HYP smearing of the Wilson link. The smearing of the Wilson link can suppress the linear divergence, such that the exponential decay with $z$ becomes weaker which eventually improves the signal to noise ratio of the data. The results for lattice spacings $a\in[0.11,0.06]$ fm differ from each other by less than 1\% while the cases without HYP smearing have larger statistical uncertainties. Such a difference can come from systematic uncertainties caused by a mismatch of the pion mass, mixed action effects and/or a residual linear divergence which is too small to be visible in the above lattice spacing range.

 \begin{figure}[tbph]
  \centering
  \includegraphics[width=8cm]{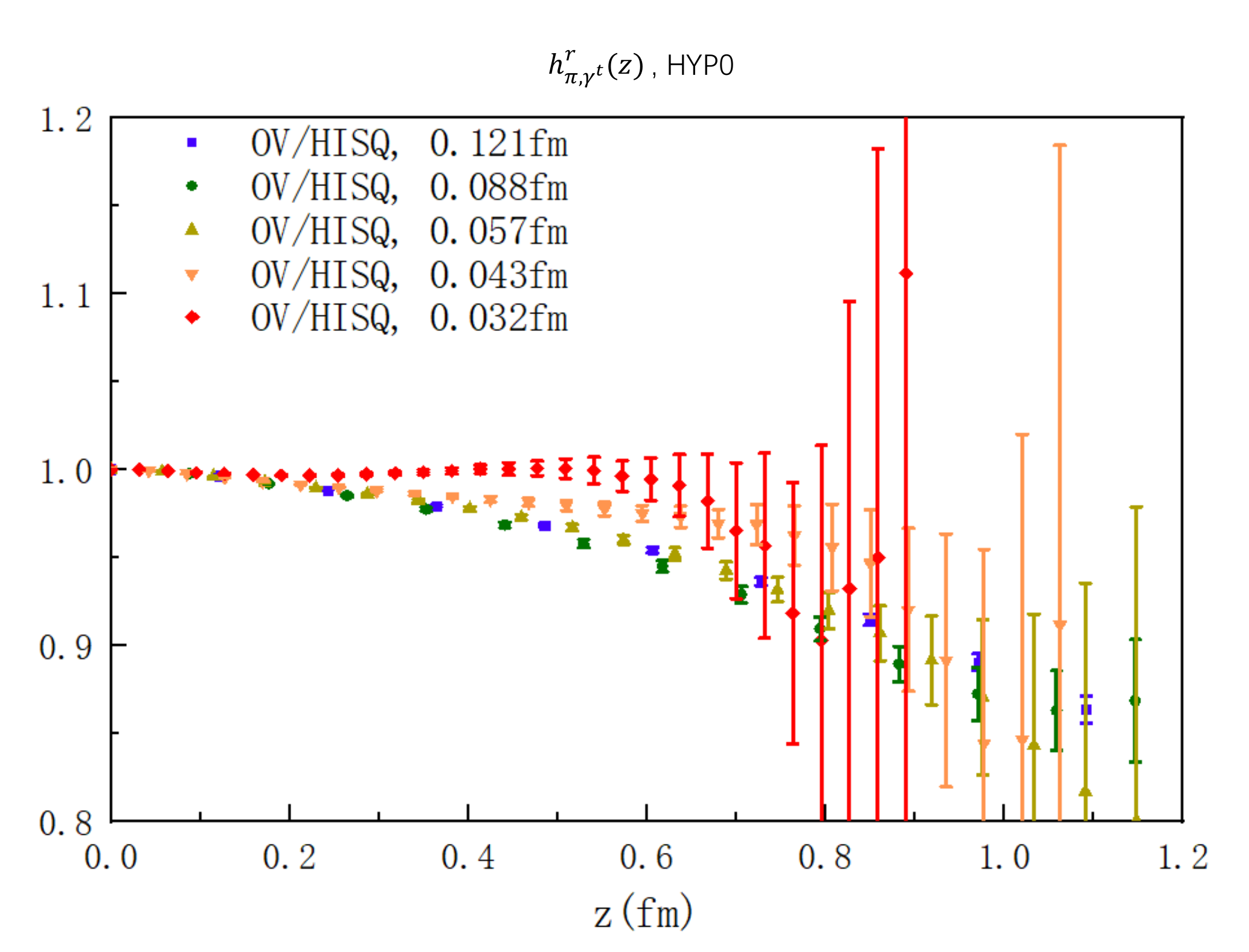}
  \includegraphics[width=8cm]{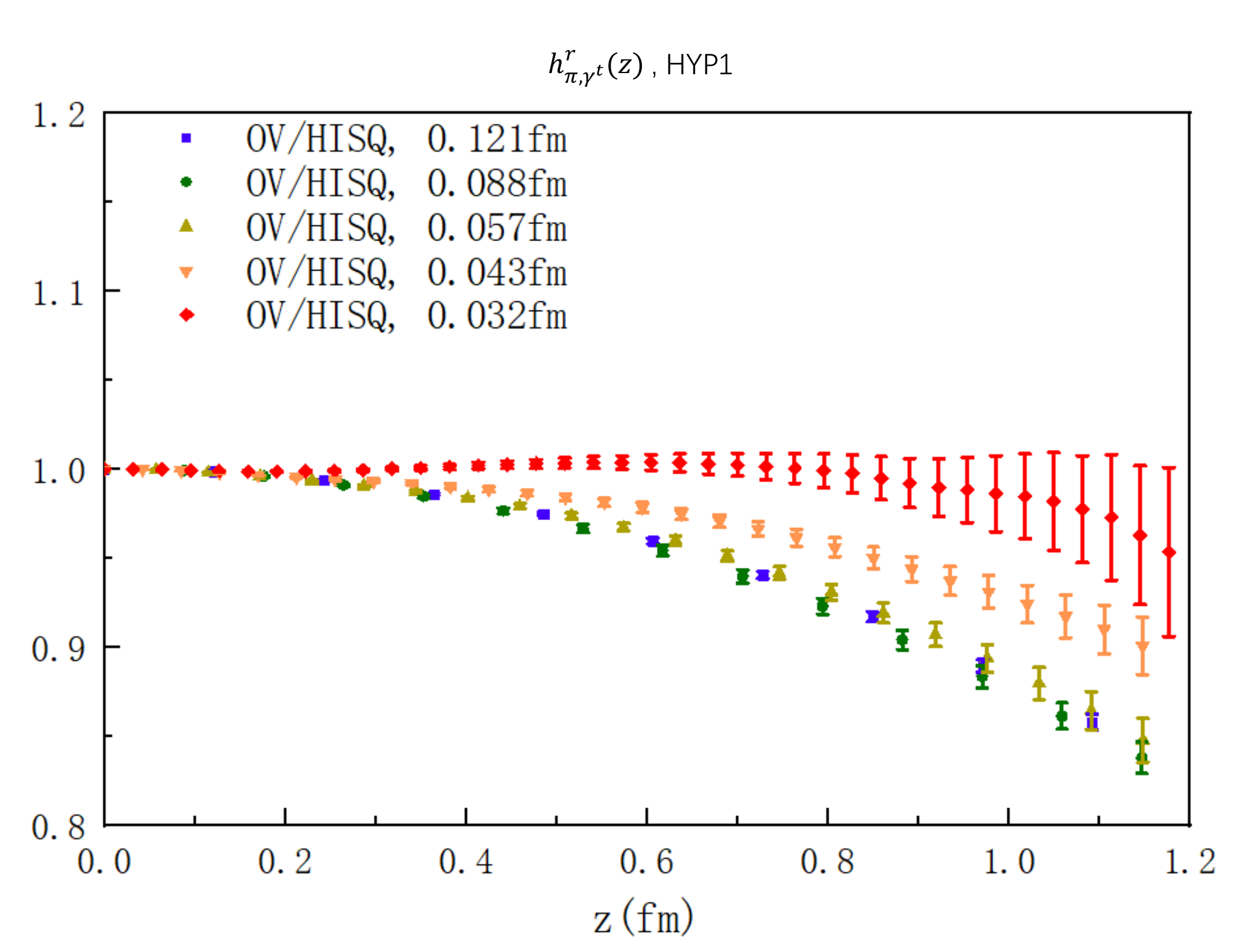}
  \caption{The RI/MOM renormalized pion matrix element of the vector current, using overlap (OV) fermions on HISQ sea ensembles and using a Wilson link without HYP smearing (upper panel) and with 1-step of HYP smearing (lower panel). The residual linear divergences in both cases are pronounced and similar.}
  \label{fig:overlap2}
  \end{figure}

Thus we switch to the other class of ensembles using the HISQ sea, which covers  lattice spacings from 0.032 fm to 0.121 fm, see Fig.~\ref{fig:overlap2}. The results for lattice spacings $a>0.05$ fm are consistent with each other, while a deviation becomes more and more obvious when the lattice spacing is reduced. The lattice spacings $(a_0,a_1,a_2)=(0.0318(3), 0.0425(4),0.576(5))$ fm form a geometrical series satisfying $a_0/a_1=a_1/a_2$ within uncertainties. The ratio $h^r_{\pi, \gamma_t}(z,a_0)/h^r_{\pi, \gamma_t}(z,a_1)$ is obviously larger than $h^r_{\pi, \gamma_t}(z,a_1)/h^r_{\pi, \gamma_t}(z,a_2)$. At the same time, both of them deviate from 1 exponentially with increasing length of the Wilson link $z$. Thus, the residual lattice spacing dependence in $h^r_{\pi, \gamma_t}(z,a)$ is likely to be caused by a linearly divergent term proportional to  $1/a$, not a logarithmic $\mathrm{log}(a)$ one.

\subsection{RI/MOM renormalized pion matrix element with the Clover action}

  \begin{figure}[tbp]
  \centering
  \includegraphics[width=8cm]{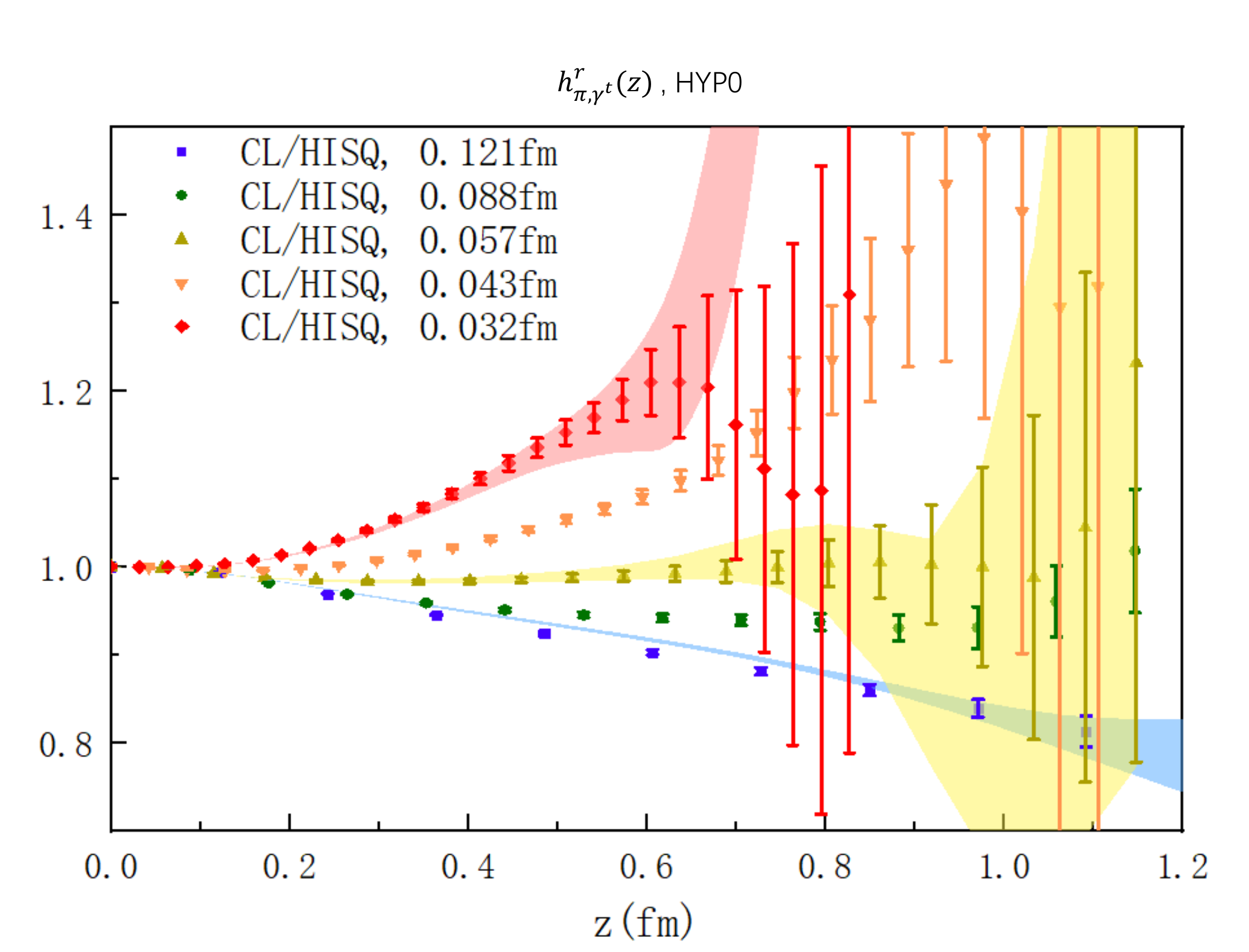}
   \includegraphics[width=8cm]{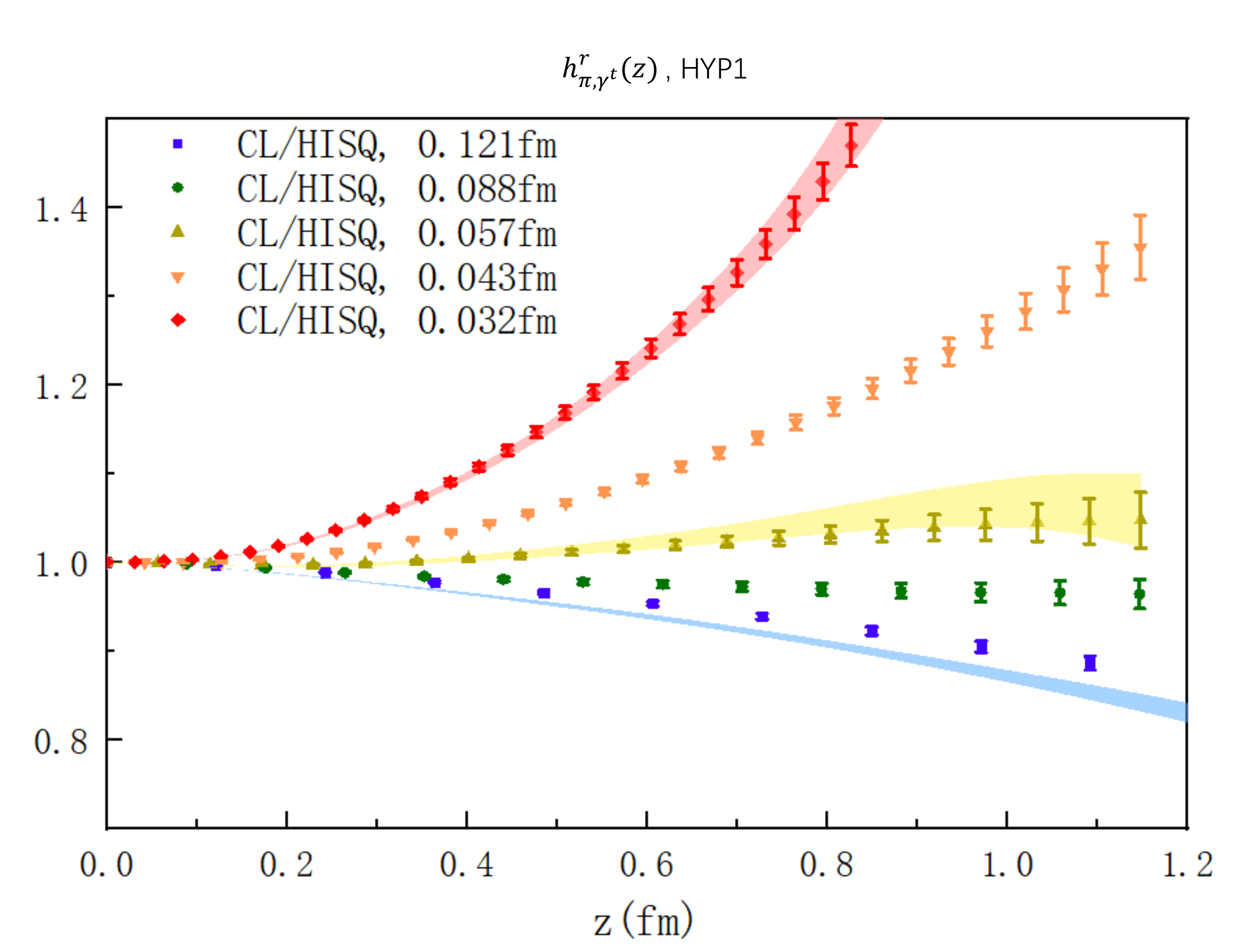}
  \caption{The same RI/MOM renormalized pion matrix element as in Fig~\ref{fig:overlap2}, but using clover (CL) fermions on HISQ sea ensembles. The residual linear divergences are much more obvious than for the overlap case. The colored bands show the results when the clover coefficient are tuned such that  the critical quark mass $|m^{\rm cri}|\le 0.02$. The residual linear divergences are still roughly the same.}
  \label{fig:clover}
  \end{figure}

  In the clover fermion case, the RI/MOM renormalized $h^r_{\pi,\gamma_t}(z,a)$ can have a very strong lattice spacing dependence.
  Since the lattice spacing dependence becomes systematically stronger for smaller lattice spacings and increases exponentially with $z$, it is probably also due to a residual linear divergence term. We also modified the clover coefficient by $\sim$ 20\% to reduce  the critical quark mass to $|m^{\rm cri}|\le0.02$. However, the lattice spacing dependence is not affected by this tuning  (see the colored bands in Fig.~\ref{fig:clover}). We also repeated the calculation with twisted-mass fermions on MILC ensembles at $a$=0.057 fm. The result agrees with that for clover fermions and obviously differs from that for  overlap ones. 
 
   \begin{figure}[tbp]
  \centering
   \includegraphics[width=8cm]{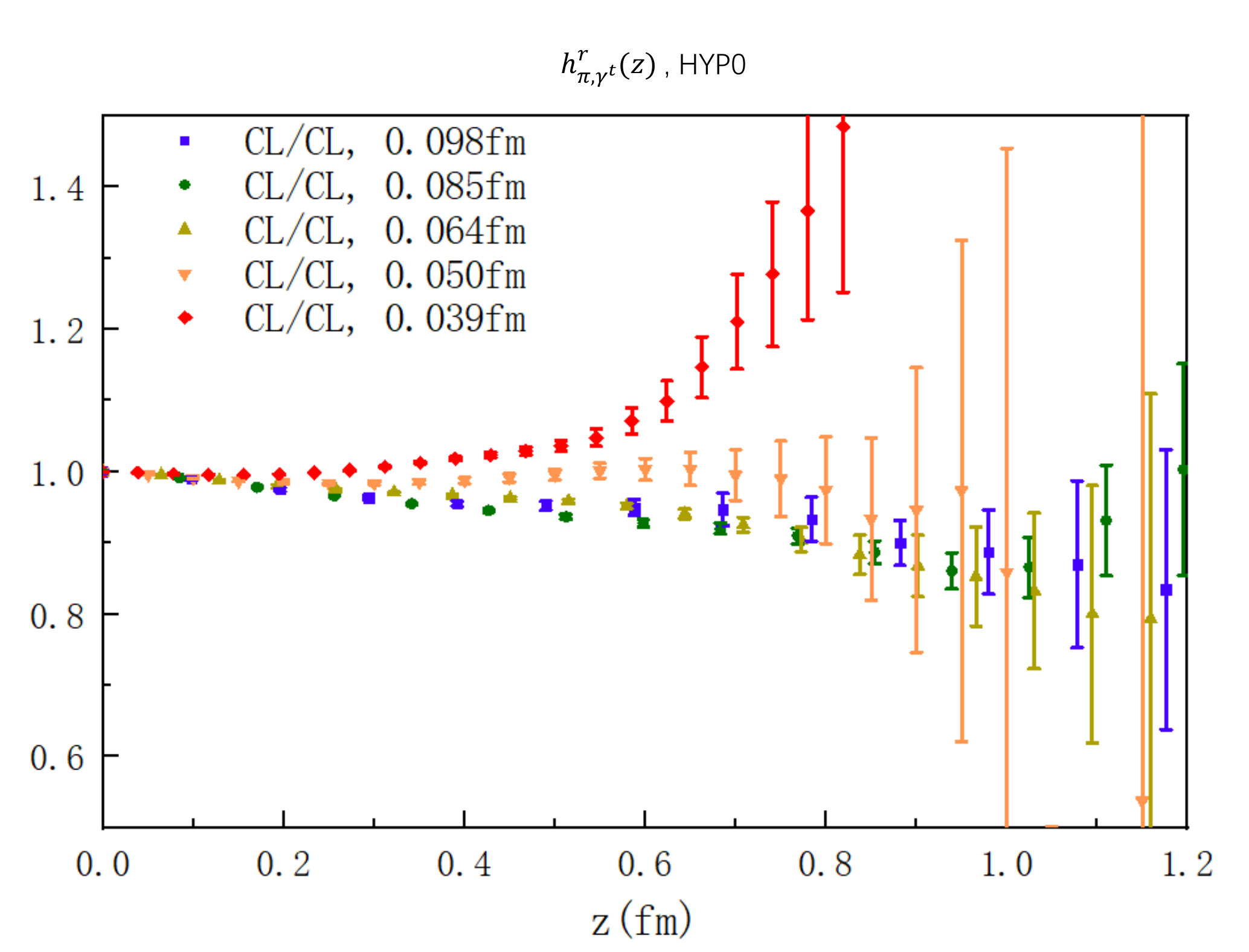}
  \includegraphics[width=8cm]{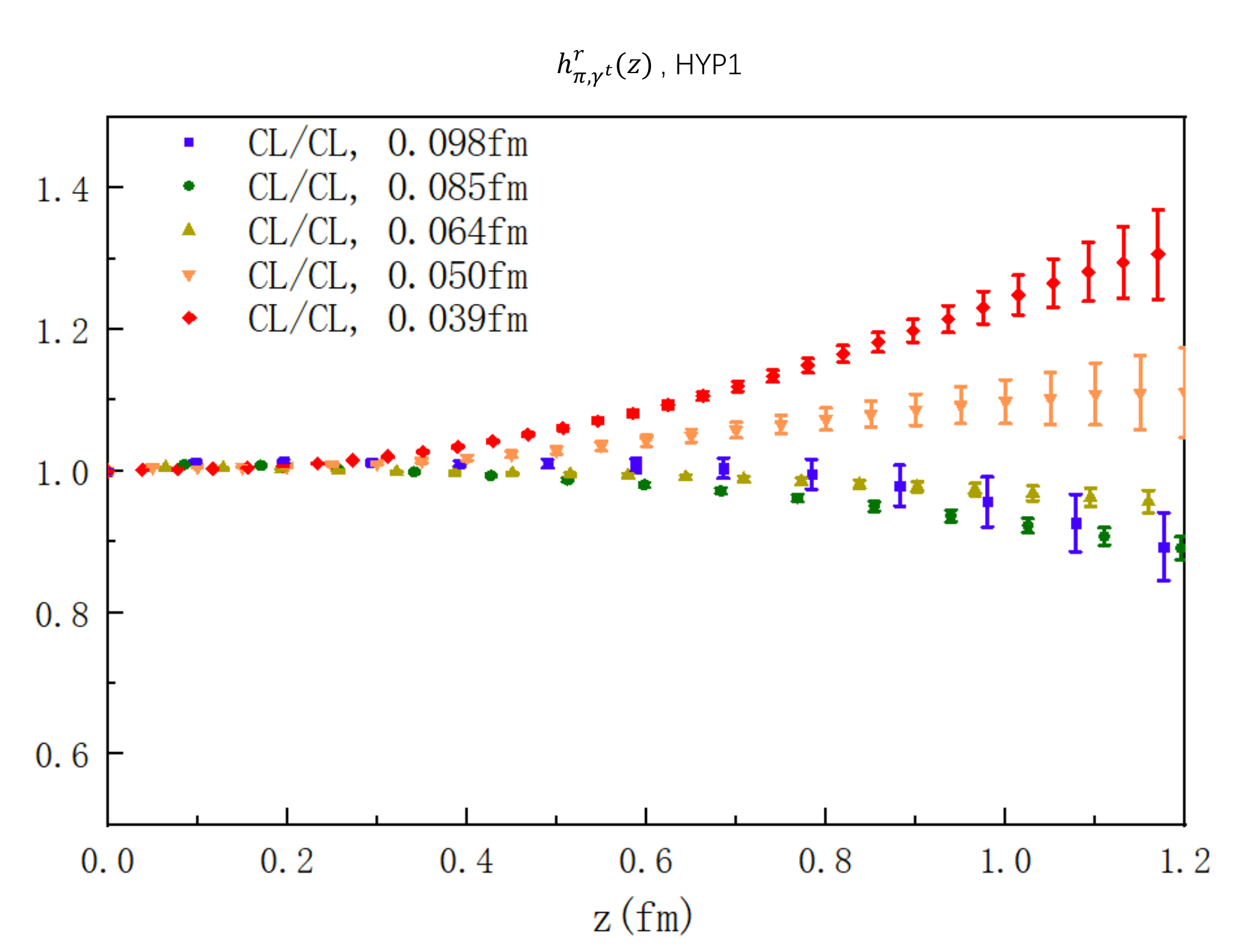}
  \caption{The same RI/MOM renormalized pion matrix element as in Fig~\ref{fig:clover}, but using clover (CL) fermions on CL sea ensembles. The residual linear divergence is still there}
  \label{fig:clover2}
  \end{figure}

 Since $m^{\textrm{cri}}_{q}$ is very sensitive to whether the gauge configuration used in the clover fermion action is HYP smeared, and the clover on HISQ setup can suffer from ${\cal O}(a)$ mixed action effects, we also studied the unitary clover fermion case using CLS ensembles. In the upper panel of  Fig.~\ref{fig:clover2} we show the results without any HYP smearing for either the fermion action or Wilson link. The situation improved somewhat at larger lattice spacing but the results at given $z$ still increase with $1/a$ except for the largest lattice spacing where discretization error are probably large. We also repeated the calculation with HYP smeared Wilson links, and found the residual linear divergence to be much more obvious because  the signal has smaller error bars.
 
It is natural to guess that the residual linear divergence is stronger in the clover fermion case than in the overlap case because of  chiral symmetry breaking effects. 
Since the operator product expansion of ${\cal O}_{\gamma_t}(z,a)$ can include trace terms like $\bar{\psi}\gamma_t D^{2}\psi$ and additional linear divergences from the Wilson, clover and/or residual mass terms, there are even several effects which can add up to produce a significant residual linear divergence.

\section{Possible origin of the residual linear divergence}

In Ref.~\cite{Huo:2021rpe}, the Lattice Parton Collaboration parameterized the linear divergence as
\begin{align}
{\rm ln}M(z,a)=\frac{kz}{a {\rm ln}[a\Lambda_{QCD}]}~+~...~,
\end{align}
where $M$ can be either the RI/MOM renormalization constant $Z$ or the pion matrix element $h$, and used our data to extract the linear divergence coefficients $k$ in different cases.  As in Table IV-V of Ref.~\cite{Huo:2021rpe} for the HYP smeared Wilson link, the $k$ is in the range k$\in[0.49,0.52]$ for the bare pion matrix elements with different valence and sea actions using $\Lambda_{QCD}=0.39$ GeV, but the $k$ in $Z$ is 0.55 for overlap fermions and $\sim$0.63 for clover fermions, on  HISQ sea ensembles. The value of k in the RI/MOM renormalized pion matrix element is $\sim$ 0.05 and 0.13 for the overlap and clover fermion cases respectively, and much smaller than that in the bare pion matrix elements. 

According to the lattice perturbative theory calculation~\cite{Constantinou:2017sej}, the linear divergence just comes from the Wilson link and is independent of the external quark state at the 1-loop level. This suggests that if there is any external state (like a quark state using different actions or the pion state) dependence in the quasi-PDF matrix element, it can only come from 2-loop or even higher level contributions, which is consistent with our observation as the residual linear divergence left after the cancelation is only at the 20\% level.

\begin{figure}[tbph]
  \centering
  \includegraphics[width=8cm]{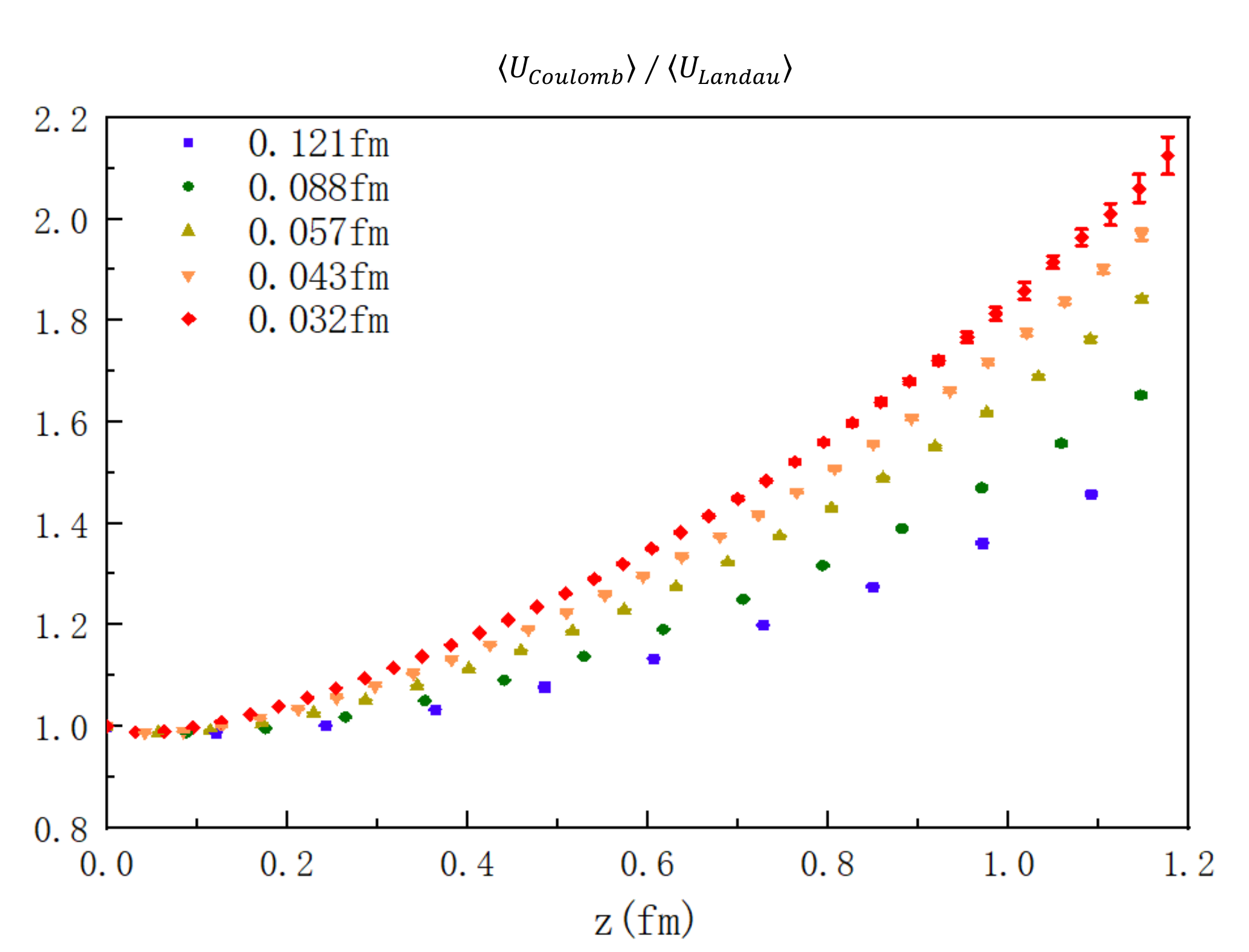}
  \caption{The ratio of gauge links in Coulomb gauge and Landau gauge on MILC ensembles with 1-step of the HYP smearing. These curves seem to converge at small lattice spacings, which indicates that the linear divergence of the Wilson link is independent from the gauge fixing condition, at least for the Landau or Coulomb ones.}
  \label{fig:wilson_link}
  \end{figure}

We guess that the Landau gauge fixing we used for the quark matrix element introduces an additional linear divergence. As the Wilson link can have a gauge dependent logarithmic divergence at 1-loop level~\cite{Constantinou:2017sej}, we would expect it to introduce sub-leading linear divergences at higher loop levels.  Thus we considered the ratio between the Wilson link in Coulomb and Landau gauge, and show the results in Fig.~\ref{fig:wilson_link}. The curves seem to converge at small lattice spacings, which suggests that the linear divergence of the Wilson link itself is unlikely to be gauge dependent, at least for Coulomb and Landau gauge fixing. 

\begin{figure}[tbph]
  \centering
  \includegraphics[width=8cm]{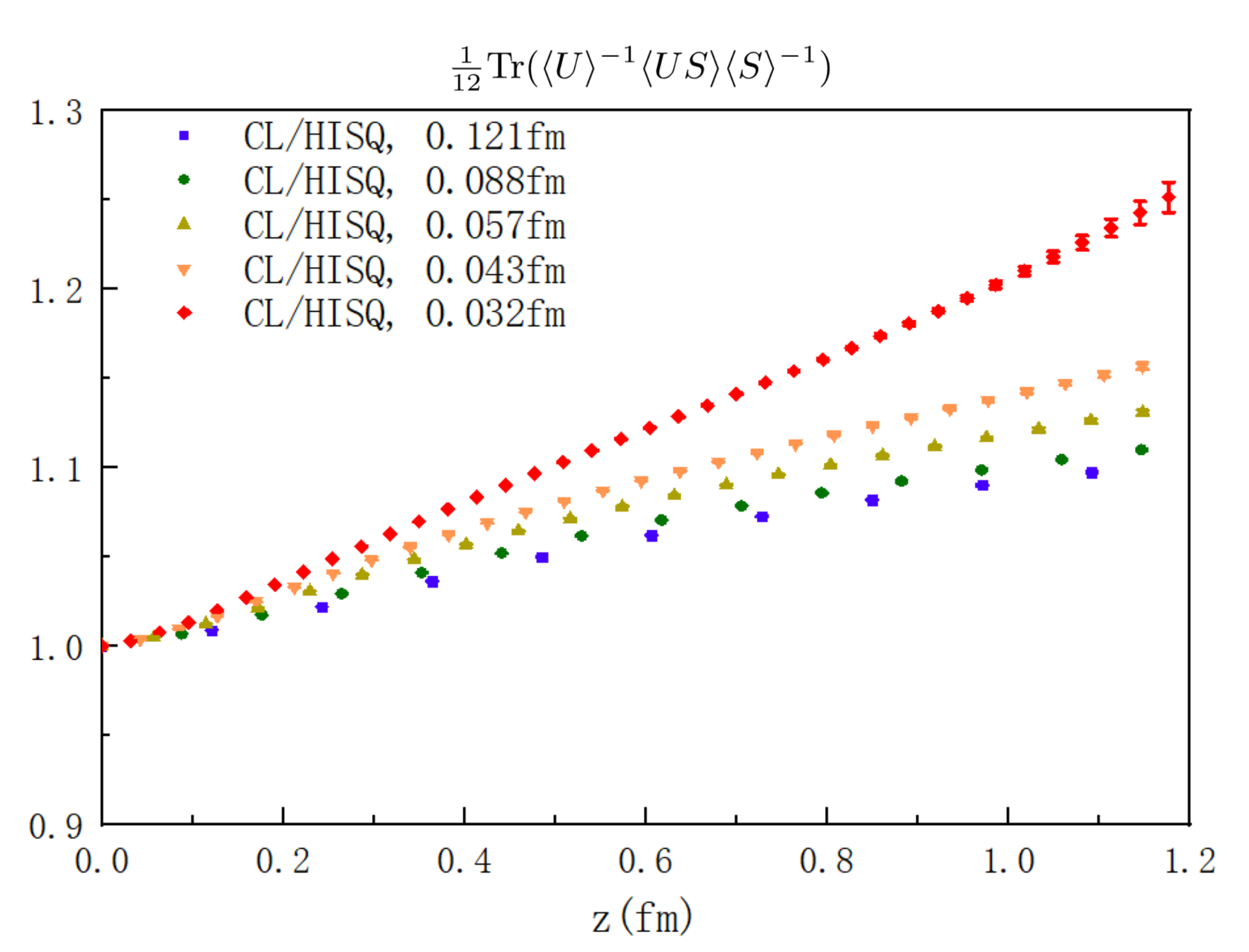}
  \includegraphics[width=8cm]{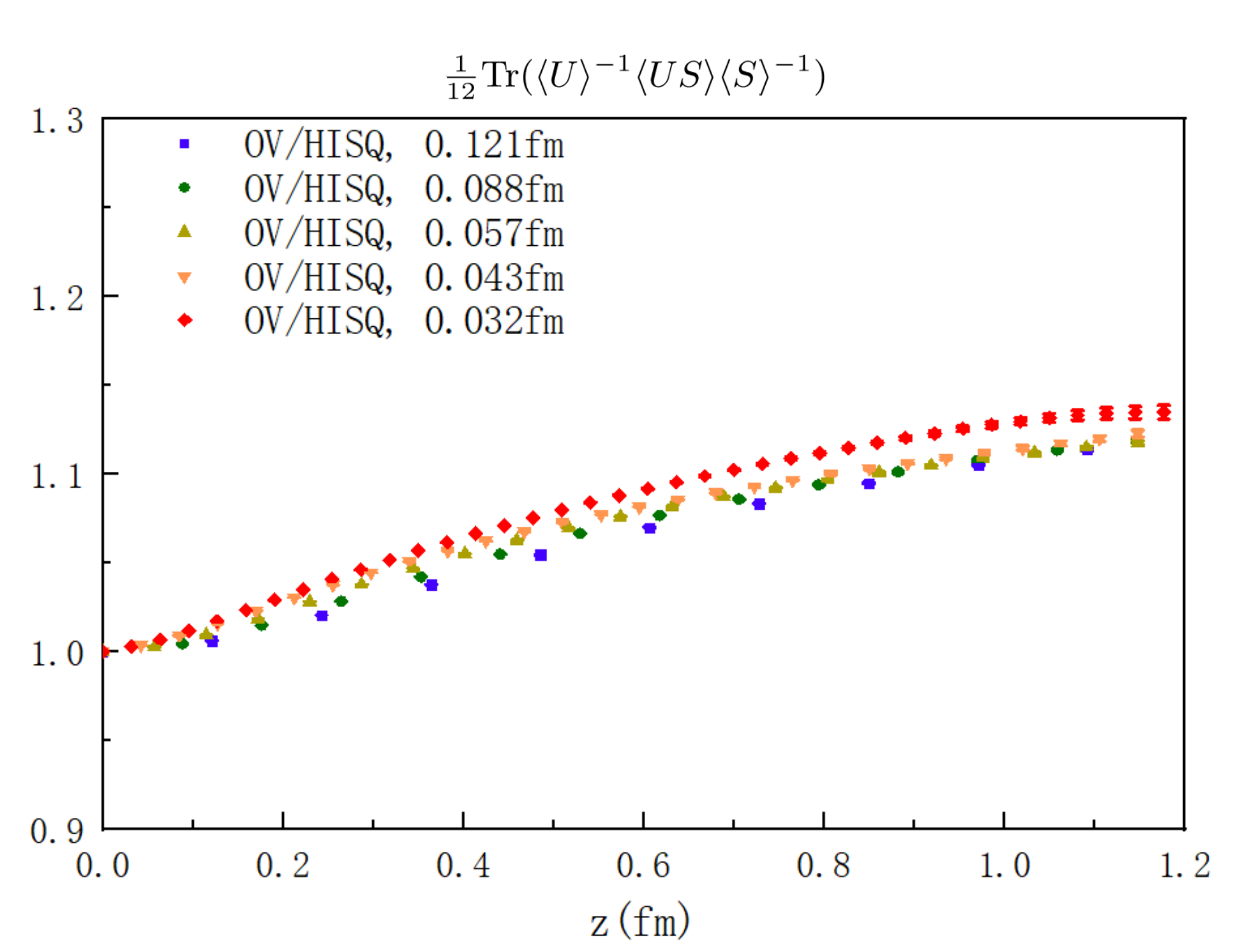}
  \caption{The value of $C_{US}(z)$ defined in Eq.~(\ref{eq:def_su}) for the subtracted correlation between the 1-step HYP smeared Wilson link and the clover fermion (upper panel) and overlap fermion (lower panel) external states, on MILC ensembles in Landau gauge. The linear divergence is obvious for both fermion actions while that in the overlap fermion case is much smaller. }
  \label{fig:su}
  \end{figure}

We also checked the correlation between the Wilson line and external states, by considering the following correlation
\begin{align}\label{eq:def_su}
C_{US}(z)=\frac{1}{12} \textrm{Tr}((\langle U(z,0)\rangle^{-1}\langle U(z,0)S(p)\rangle \langle S(p)\rangle^{-1}),
\end{align}
 where $S(p)=\sum_x S(0,x)e^{ipx}$ is the momentum projected quark propagator. In the practical calculation we averaged  over the reference points of all three expectation values in the right hand side of Eq.~(\ref{eq:def_su}) independently to improve statistics. If the linear divergence only comes from the Wilson link and is independent of the external state, then $C_{US}(z)$ has to be free of any linear divergence, as confirmed by the 1-loop lattice perturbative theory calculation~\cite{Constantinou:2017sej}. However, as shown in Fig.~\ref{fig:su} for the clover (upper panel) and overlap fermion (lower panel) on HISQ sea ensembles with 1-step of the HYP smearing of the link, we find the linear divergences to exist in both, the clover and overlap case. It is, however, much smaller in the latter. This is similar to what we saw for the RI/MOM renormalized pion quasi-PDF matrix element. However, since a similar test cannot be done for the Coulomb gauge due to the residual gauge degree of freedom along the temporal direction, we cannot yet verify whether the residual linear divergence we saw in $h^r_{\pi,\gamma_t}(z)$ and $C_{US}(z)$ are gauge dependent or not.

\begin{figure}[tbph]
  \centering
  \includegraphics[width=8cm]{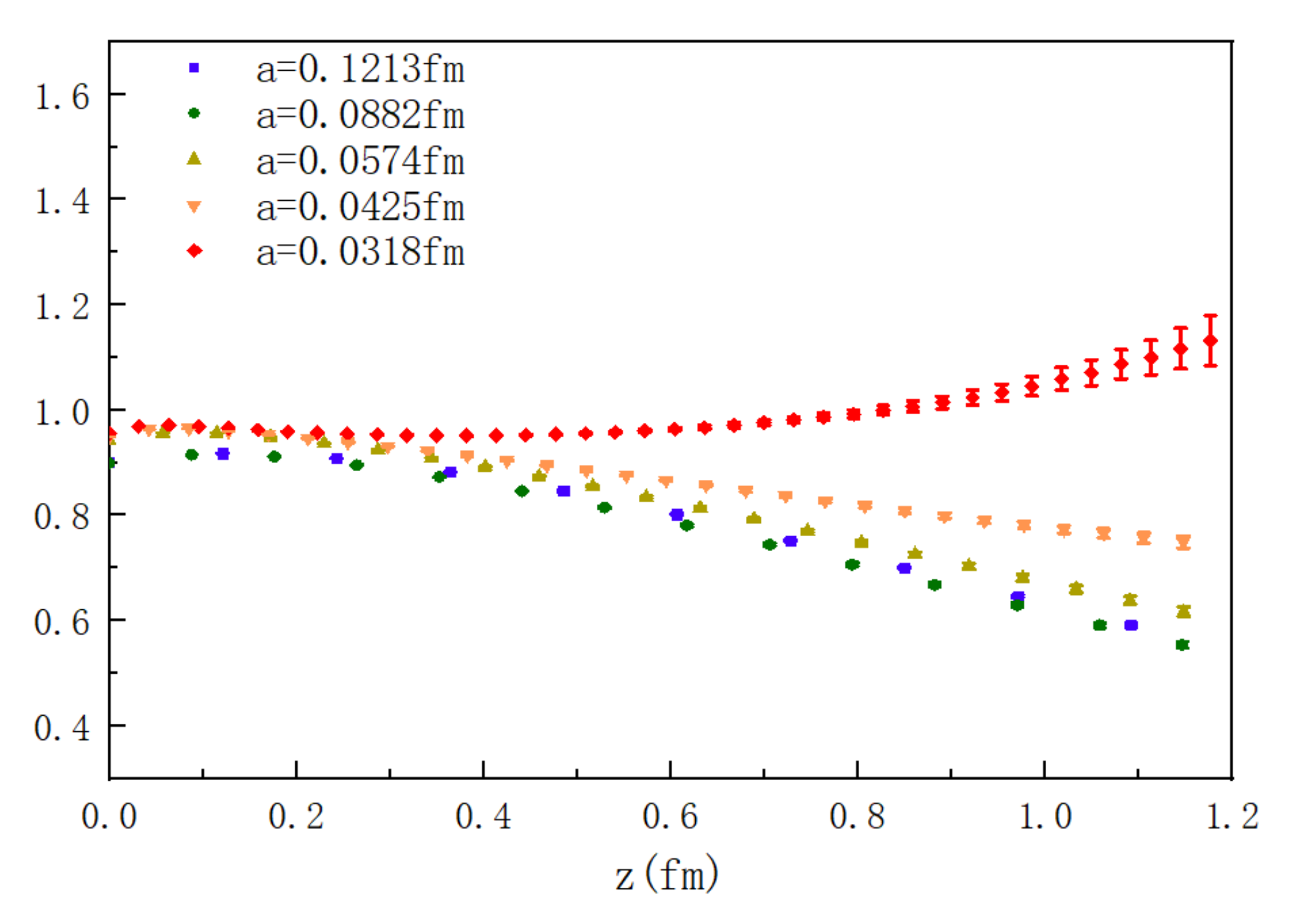}
  \caption{The pion quasi-PDF matrix element renormalized by the Coulomb gauge fixed Wilson line, using the overlap fermion on HISQ ensembles. }
  \label{fig:link_renorm}
  \end{figure}

Eventually we consider the ratio
 \begin{align}\label{eq:renorm_u}
h^{r,U}_{\pi,\gamma_t}(z)=\frac{h_{\pi,\gamma_t}(z)}{\frac{1}{3}\textrm{Tr}\langle U\rangle},
\end{align}
with the Wilson link in the Coulomb gauge, to test whether the linear divergence for the hadron matrix element is exactly the same as that for the Wilson link. As seen in Fig.~\ref{fig:link_renorm}, $h^{r,U}_{\pi,\gamma_t}(z)$ also has a similar lattice spacing and $z$ dependence pattern as we observed  for $h^{r}_{\pi,\gamma_t}(z)$ and $C_{US}(z)$. This suggests that the linear divergence for the pion quasi-PDF matrix element is also not exactly the same as that of the Wilson line itself. Since we have shown that the linear divergence in the pion matrix element is independent of the fermion action (Fig.~\ref{fig:ratio1}), and that in the Wilson line it is insensitive to whether one chooses Landau or Coulomb gauge fixing (Fig.~\ref{fig:wilson_link}), we will skip the discussion on the other combinations. 

 \section{Summary}

 In this contribution we study systematically the continuum limit for quasi-PDFs with RI/MOM renormalization in the LaMET approach. We compare results for a variety of valence and sea quark fermion actions and lattice set-ups. We find that the traditional RI/MOM method cannot eliminate the characteristic linear divergences of quasi-PDFs completely, and that the remnants of this linear divergence blow up with decreasing lattice spacing. Obviously this greatly complicates control of the continuum limit.

We make a number of observations which can prove useful to clarify the origin of these remnants and develop strategies to cope with them:\\ 

The RI/MOM renormalized pion matrix elements in the rest frame include residual linear divergences from certain higher loop effects, so, higher loop calculation in lattice perturbation theory are needed.\\

Simulations with chiral fermions (overlap or domain-wall) are much less affected than those with clover fermions, so, chiral symmetry seems to play an important role.\\

Since the renormalization of operators should be frame independent, we believe that this conclusion should also apply to all quasi-PDF calculations in a moving frame. For a quasi-PDF calculation using an ensemble with $a=0.09$ fm and clover fermions, we estimate the systematic bias to be $\sim 5$\% based on the correction at $z\sim$ 1.0 fm which is smaller than kinds of the systematic uncertainties in the practical quasi-PDF calculation; but it should increase to more than 50\% at $a=0.03$ fm. Thus, removal of the residual linear divergence is crucial to obtain a meaningful continuum limit. For overlap fermions the corresponding bias could be 10\%  at 0.03 fm which might still be acceptable compared to other sources of uncertainty, but the fundamental problem is the same.\\ 

We believe that similar checks should be done for all LaMET calculations using bi-linear operators with Wilson link, e.g.,  quasi-PDFs, quasi-DAs, quasi-TMDs and so on. Comparison to calculations with large momentum for the nucleon suggests that the proposed check just requires the calculation of ${\cal O}(200)$ propagators on each ensemble, while providing essential information to avoid possible misinterpretations of the results obtained, which is obviously a very well justified investment. For the pseudo-PDFs for which a ratio of hadron matrix elements is used for renormalization, it is also essential to verify that the linear divergence is indeed completely canceled, independent of all details of the specific simulation, e.g., of the momenta used.

The outcome of many such tests should allow to better understand the origin of these residues and should provide clues for how to remove them systematically~\cite{Huo:2021rpe}. Let us stress that these residues become only sizable at small lattice spacings and that it is highly probable that the results obtained for not so fine lattices extrapolate smoothly to the continuum. So, existing results stay valid. However, the ultimate goal must be full control of the continuum limit and, therefore, this problem has to get solved.  

The perturbative and non-perturbative study in lattice regularization beyond the 1-loop level is also essential, as a residual linear divergence is forbidden at the 1-loop level. We also confirmed the following expectations:

1) The linear divergences in the pion quasi-PDF matrix element with different fermion actions (overlap and clover) are the same. Combining with the consistency of the pion and nucleon matrix element shown in Ref.~\cite{Huo:2021rpe}, we conclude that the linear divergences in different hadrons are the same, regardless of the fermions we use.

2) The linear divergence in the Wilson link is indeed gauge independent up to certain lattice artifacts, based on our calculation in the Coulomb and Landau gauges.

3) There is a residual linear divergence due to the gluon exchange between the external state and the Wilson line, which is absent at 1-loop level. Such a linear divergence is action-dependent for a quark state in the Landau gauge, while it is action-independent in the hadron state.

\section*{Acknowledgement}
We thank the CLS,  MILC and RBC/UKQCD collaborations for providing us their gauge configurations, and Long-Cheng Gui, Xiangdong Ji, Keh-Fei Liu, 
Wei Wang, Jian-Hui Zhang and Yong Zhao for useful information and discussion. The calculations were performed using the Chroma software suite~\cite{Edwards:2004sx} with QUDA~\cite{Clark:2009wm,Babich:2011np,Clark:2016rdz} and GWU-code~\cite{Alexandru:2011ee,Alexandru:2011sc} through HIP programming model~\cite{Bi:2020wpt}.
The numerical calculation has majorly been done on CAS Xiaodao-1 computing environment, and  supported by Strategic Priority Research Program of Chinese Academy of Sciences, Grant No. XDC01040100, HPC Cluster of ITP-CAS, and Jiangsu Key Lab for NSLSCS.
P. Sun is supported by Natural Science Foundation of China under grant No. 11975127, as well as Jiangsu Specially Appointed Professor Program. Y. Yang is  supported by Strategic Priority Research Program of Chinese Academy of Sciences, Grant No. XDC01040100, XDB34030303, and XDPB15. A. Sch\"afer, P. Sun and Y. Yang are also supported by a NSFC-DFG joint grant under grant No. 12061131006 and SCHA 458/22.
\bibliography{ref}


\begin{widetext}

\section*{Appendix}

\subsection{Operator mixing and projection scheme in the RI/MOM renormalization}

The expression for the amputated Green's function reads
\begin{align}\label{eq:mixing_pattern0}
\Lambda_{\gamma_t}(z,p)=\tilde{F}_t(z,p)\gamma_t+\tilde{F}_z(z,p)\{\gamma_z z\}p_t+\tilde{F}_p(z,p)\frac{p_tp\!\!\!/}{p^2}.
\end{align}
The renormalization constant we use for the main discussion, with the projector $\gamma_t$ (the minimum projection defined in \cite{Liu:2018uuj}), just includes the $\tilde{F}_{t}$ term. However, for the p-slash projector $p\!\!\!/ /p_t$ we have,
\begin{align}\label{eq:z_pslash}
Z_{p\!\!\!/}=\frac{Z_q(\mu)}{(\tilde{F}_t(z,p)+\tilde{F}_z(z,p)zp_z+\tilde{F}_p(z,p))_{p^2=-\mu^2}},
\end{align}
which includes the contributions from not only the $\tilde{F}_t$ term but also the $\tilde{F}_{z/p}$ terms. If the linear divergences in all three terms are exactly the same and independent of $p_z$, then the choice of the projector is irrelevant for the linear divergence cancellation. 
However, as shown in this contribution, the numerical verification is still valuable. With the momentum $2\pi(5,5,0,0)/L$ using the operator ${\cal O}_{\gamma_t}(z)$, the contributions from both $\tilde{F}_{z/p}$ and the $p_z$ dependence are absent since  $p_z=p_t=0$; but thanks to the rotational symmetry of the Euclidean 4-D lattice, we can consider operators like ${\cal O}_{\gamma_t}(x)$, ${\cal O}_{\gamma_x}(y)$ and ${\cal O}_{\gamma_x}(z)$, to obtain the necessary information to extract $\tilde{F}_{z/p}$. With fixed $z$ and $p^2$, we can extract $\tilde{F}_{t/z/p}$ as function of $p_z=k$ from the following conditions,
\begin{align}
\begin{array}{cccc}
\textrm{Tr}[\gamma_t\Lambda_{\gamma_t}]|_{p=(k,k,0,0)}=&\tilde{F}_{t}(0),&& \\
\textrm{Tr}[\gamma_x\Lambda_{\gamma_t}]|_{p=(k,0,0,k)}=& & &\ \frac{1}{2}\tilde{F}_{p}(0),\\
\textrm{Tr}[\gamma_z\Lambda_{\gamma_t}]|_{p=(k,0,0,k)}=& &\ zk\tilde{F}_{z}(0),&\\
\textrm{Tr}[\gamma_t\Lambda_{\gamma_t}]|_{p=(k,0,0,k)}=&\tilde{F}_{t}(k),&& \\
\textrm{Tr}[\gamma_t\Lambda_{\gamma_t}]|_{p=(0,0,k,k)}=&\tilde{F}_{t}(k)&&+\frac{1}{2}\tilde{F}_{p}(k),\\
\textrm{Tr}[\gamma_z\Lambda_{\gamma_t}]|_{p=(0,0,k,k)}=& & +zk\tilde{F}_{z}(k)&+\frac{1}{2}\tilde{F}_{p}(k).\\
\end{array}
\end{align}

\begin{figure}[tbph]
  \centering
   \includegraphics[width=8cm]{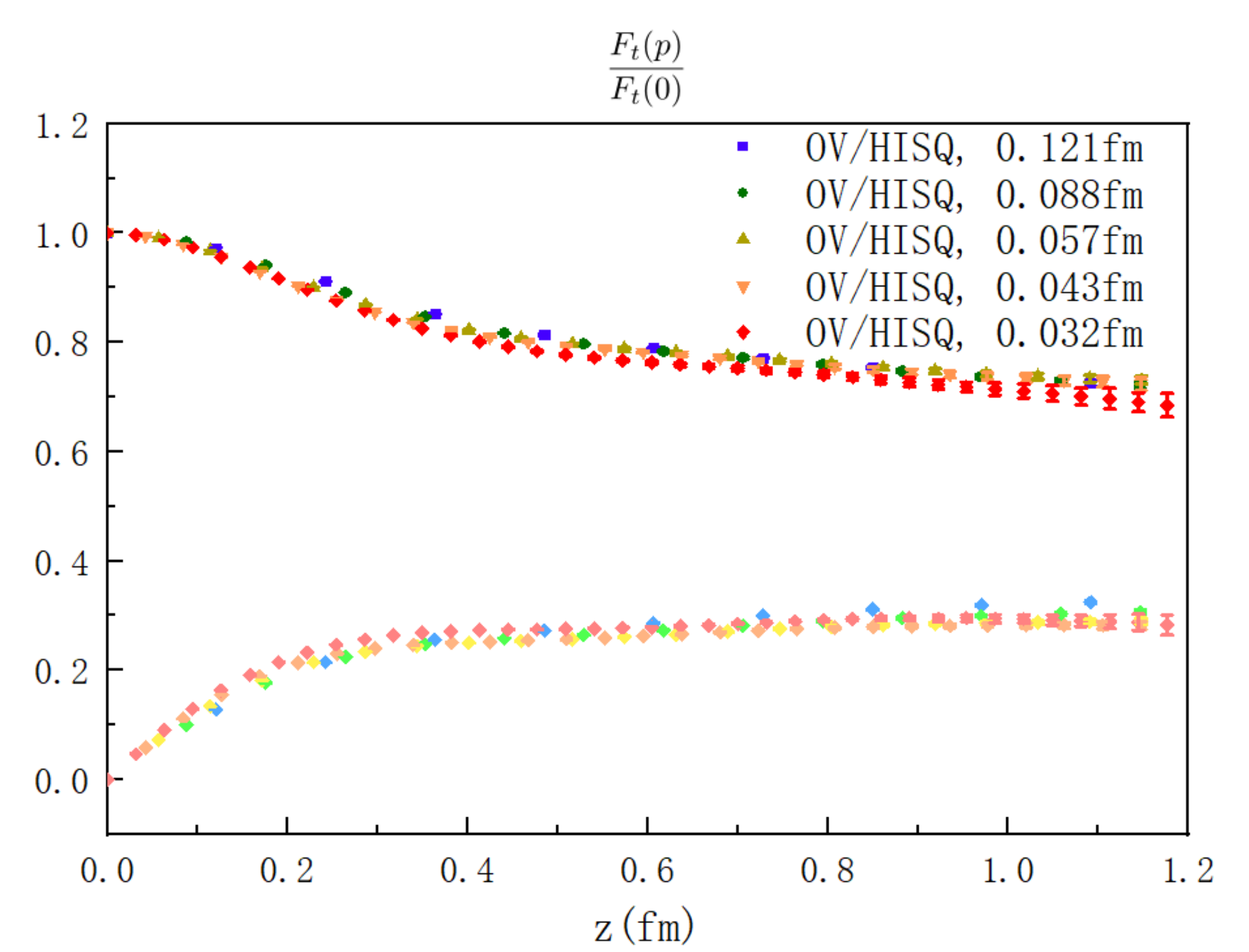}
   \includegraphics[width=8cm]{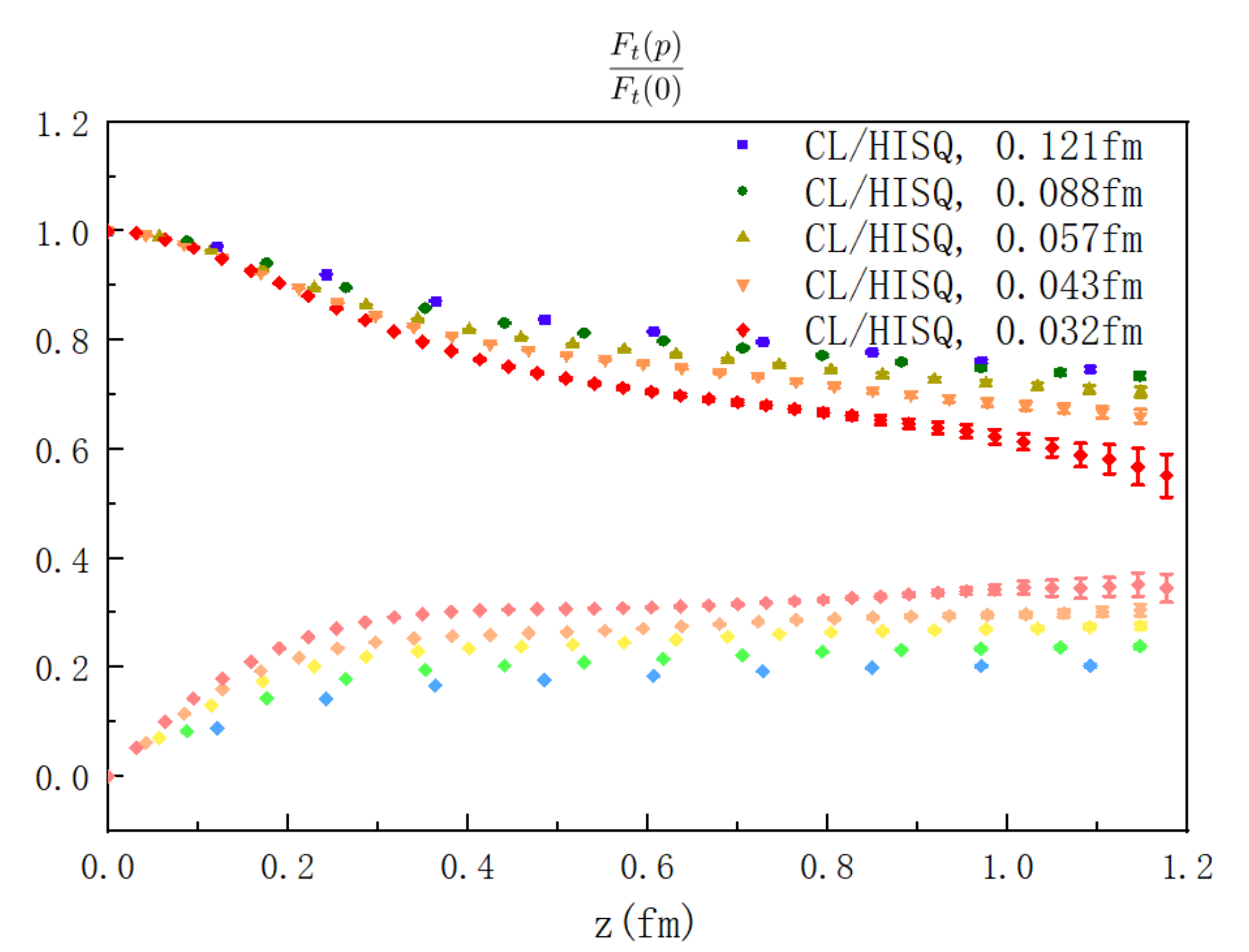}
  \caption{The $\tilde{F}_{t}(k;z)$ with non-zero momentum along the Wilson link as function of $z$ and normalized by $\tilde{F}_{t}(0;z)$, for overlap or clover fermions on HISQ sea ensembles. The darker data points show the real part and the lighter ones show the imaginary part.}
  \label{fig:mixing_terms}
\end{figure}

First of all, Fig.~\ref{fig:mixing_terms} shows the ratio $\tilde{F}_{t}(k;z)/\tilde{F}_{t}(0;z)$ using overlap and clover fermions respectively. For the overlap fermion case (left panel), the values at different lattice spacings are consistent with each other given the statistical uncertainties. This suggests that the linear divergence here is insensitive to $k$. However, in the clover case ((right panel)), it seems that there is an additional lattice spacing dependent phase angle, which makes the imaginary part to be larger at smaller lattice spacings. 

\begin{figure}[tbph]
  \centering
   \includegraphics[width=8cm]{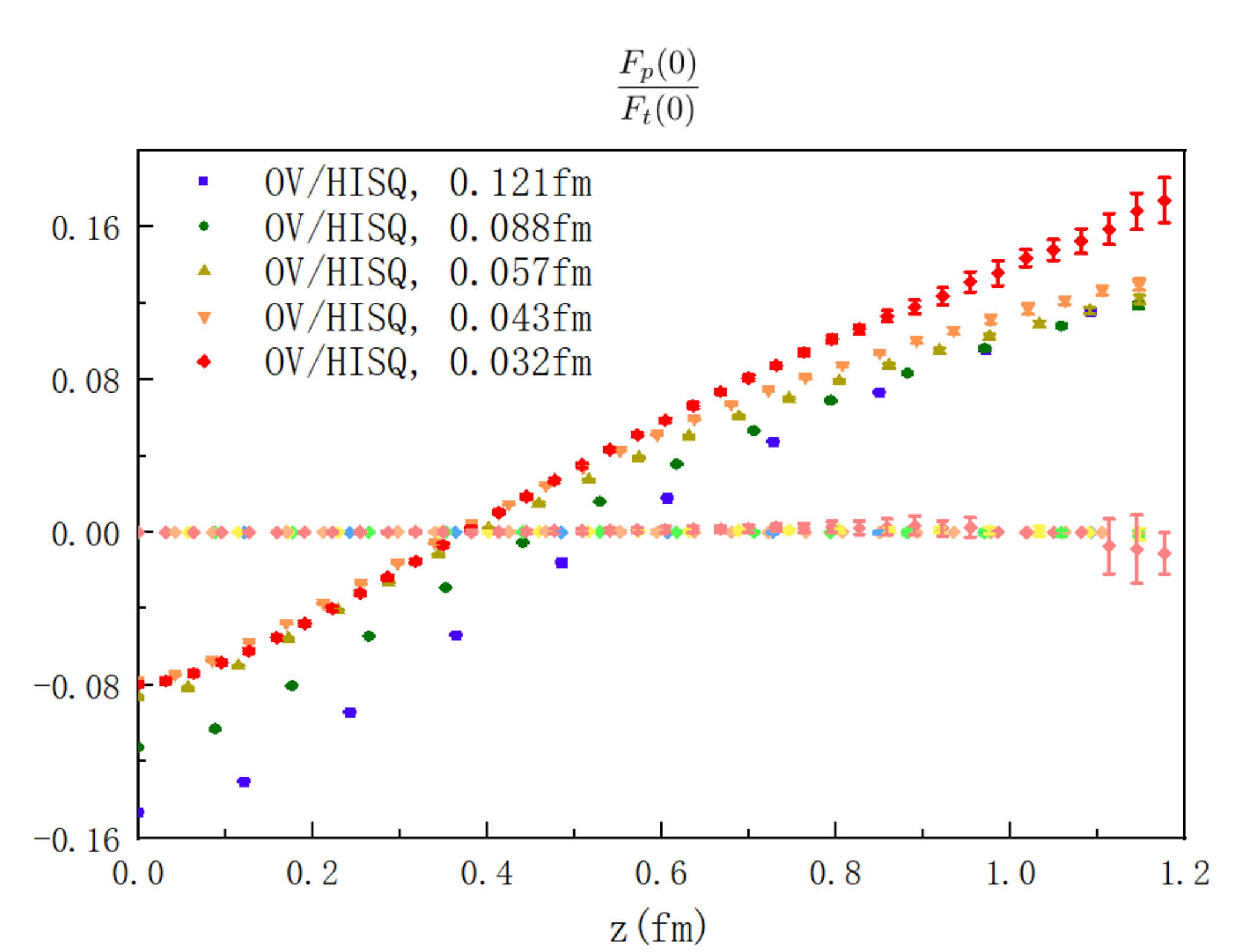}
   \includegraphics[width=8cm]{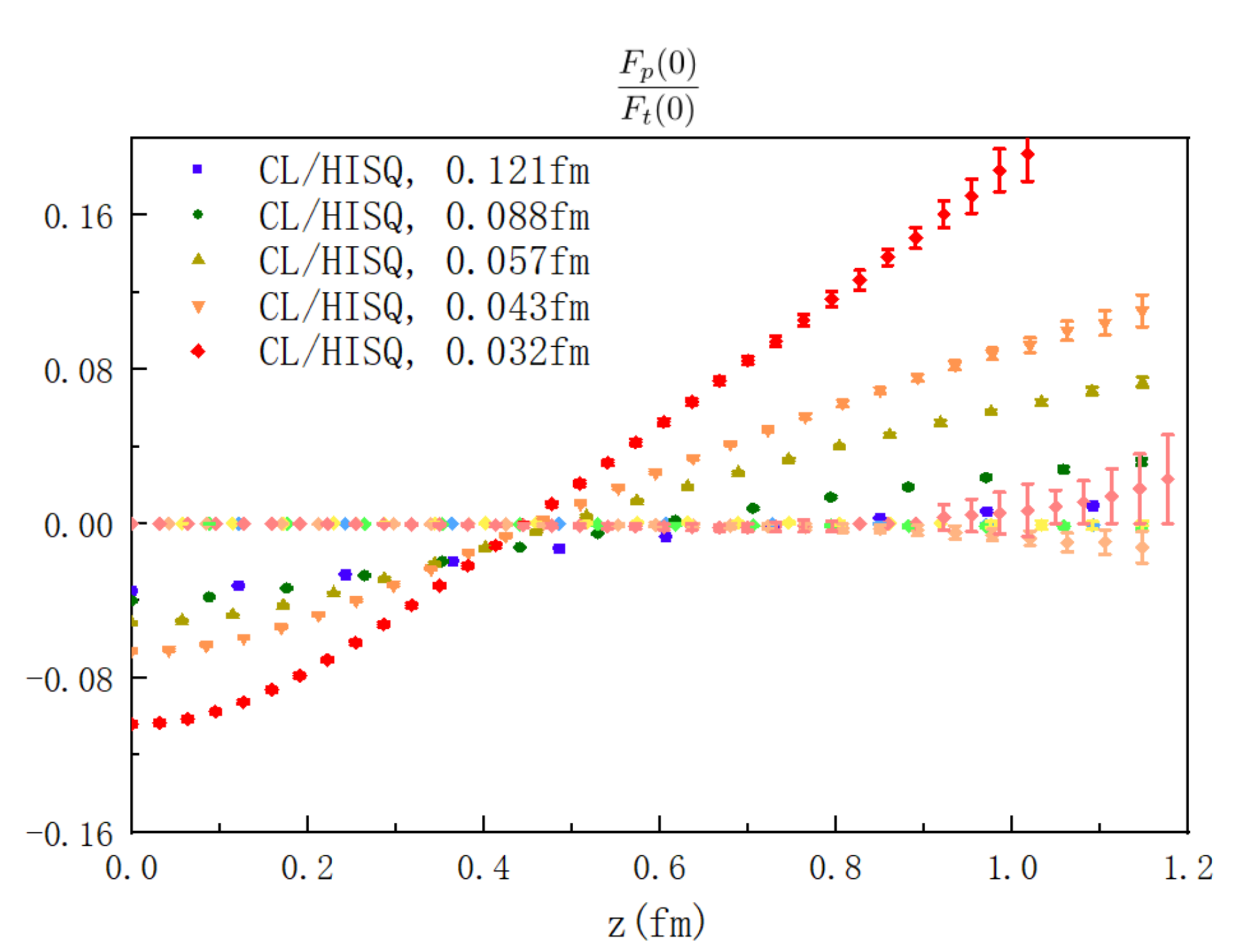}
   \includegraphics[width=8cm]{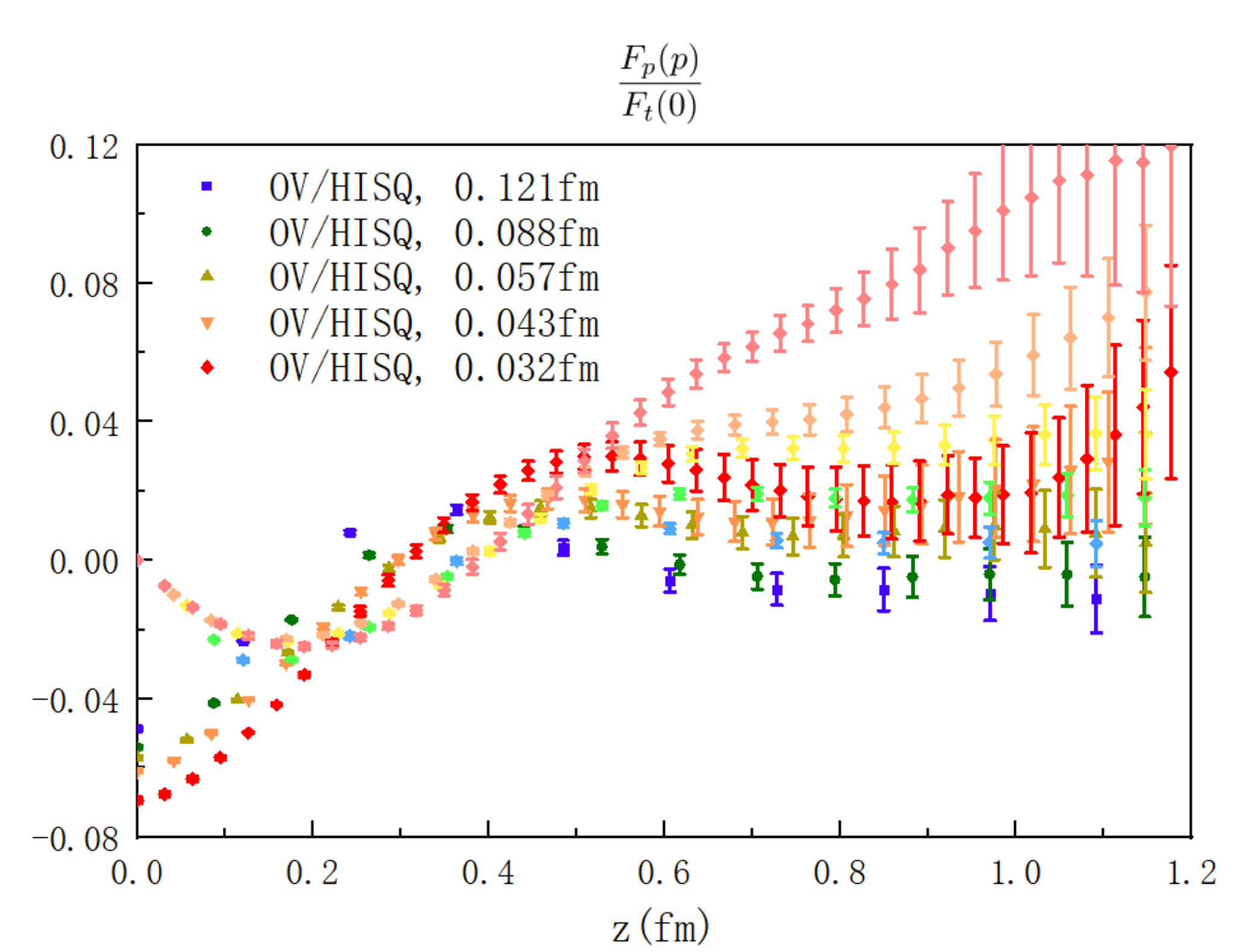}
   \includegraphics[width=8cm]{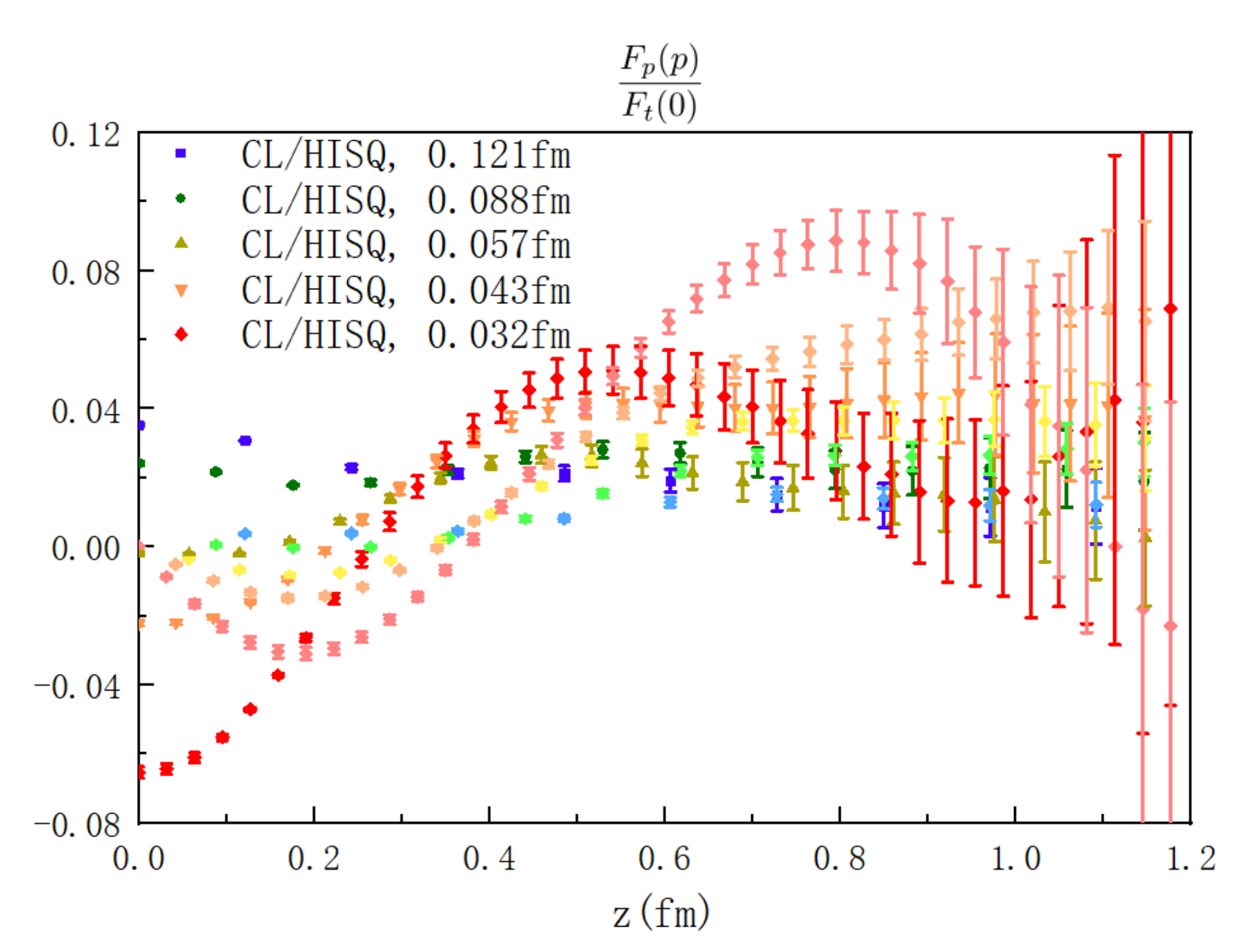}
  \caption{The $\tilde{F}_{p}(0;z)$ with zero momentum along the Wilson link (upper panels), and that with momentum $k=10\pi/L$ (lower panels), normalized by $\tilde{F}_{t}(0;z)$, using the same simulation setup as Fig.~\ref{fig:mixing_terms}.}
  \label{fig:mixing_terms2}
\end{figure}

Then we turn to the p-slash term $\tilde{F}_{p}$ with $p_z=0$. In the upper left panel of Fig.~\ref{fig:mixing_terms2}, we can see that the real part of $\tilde{F}_{p}(0;z)/\tilde{F}_{t}(0;z)$ for overlap fermions converges at small lattice spacing when $z\le 0.4$ fm, while discretization errors lead to deviations at larger lattice spacings. At large z, it seems that the discretization error becomes much smaller and then the values for different lattice spacings approach one another, while for the smallest lattice spacing there are obvious deviations. Since $\tilde{F}_{p}(0;z)$ is more than an order of magnitude smaller than $\tilde{F}_{t}(0;z)$, it is understandable that discretization errors can have a larger impact and that its competition with the linear divergence can generate a complicated lattice spacing dependence. Comparing to the overlap fermion case, the clover fermion case shown in the upper right panel can be explained similarly: the linear divergence effect in the clover case is much larger as suggested by the lattice spacing dependence at large $z$, and, therefore, the remnant linear divergent pattern is also visible at small $z$.

The cases with $p_z\neq 0$ are much more complicated, for both overlap and clover fermions as is shown in the lower two panels of Fig.~\ref{fig:mixing_terms2}. The same arguments
about the competition between the linear divergence and discretization errors also apply here. 

\begin{figure}[tbph]
  \centering
   \includegraphics[width=8cm]{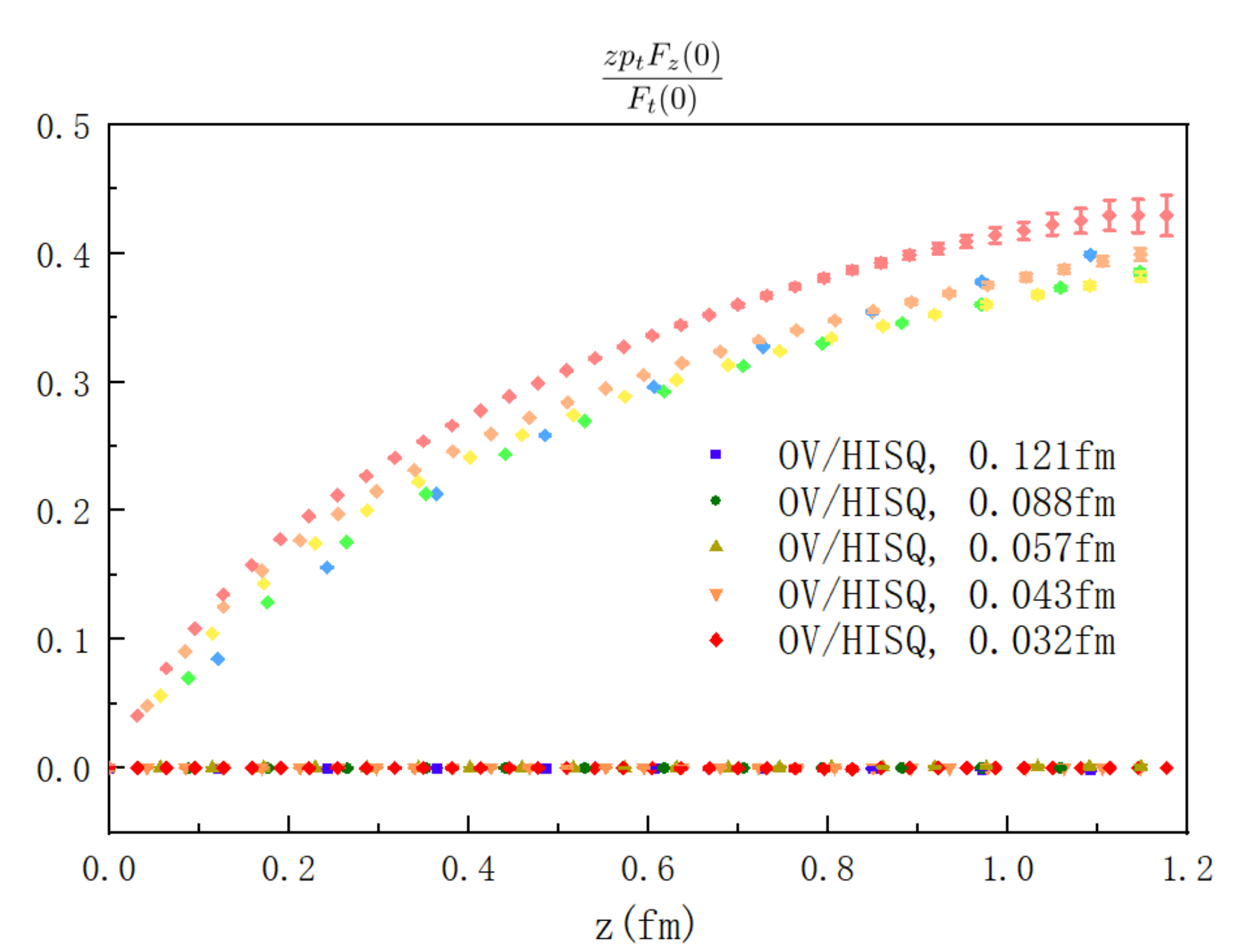}
   \includegraphics[width=8cm]{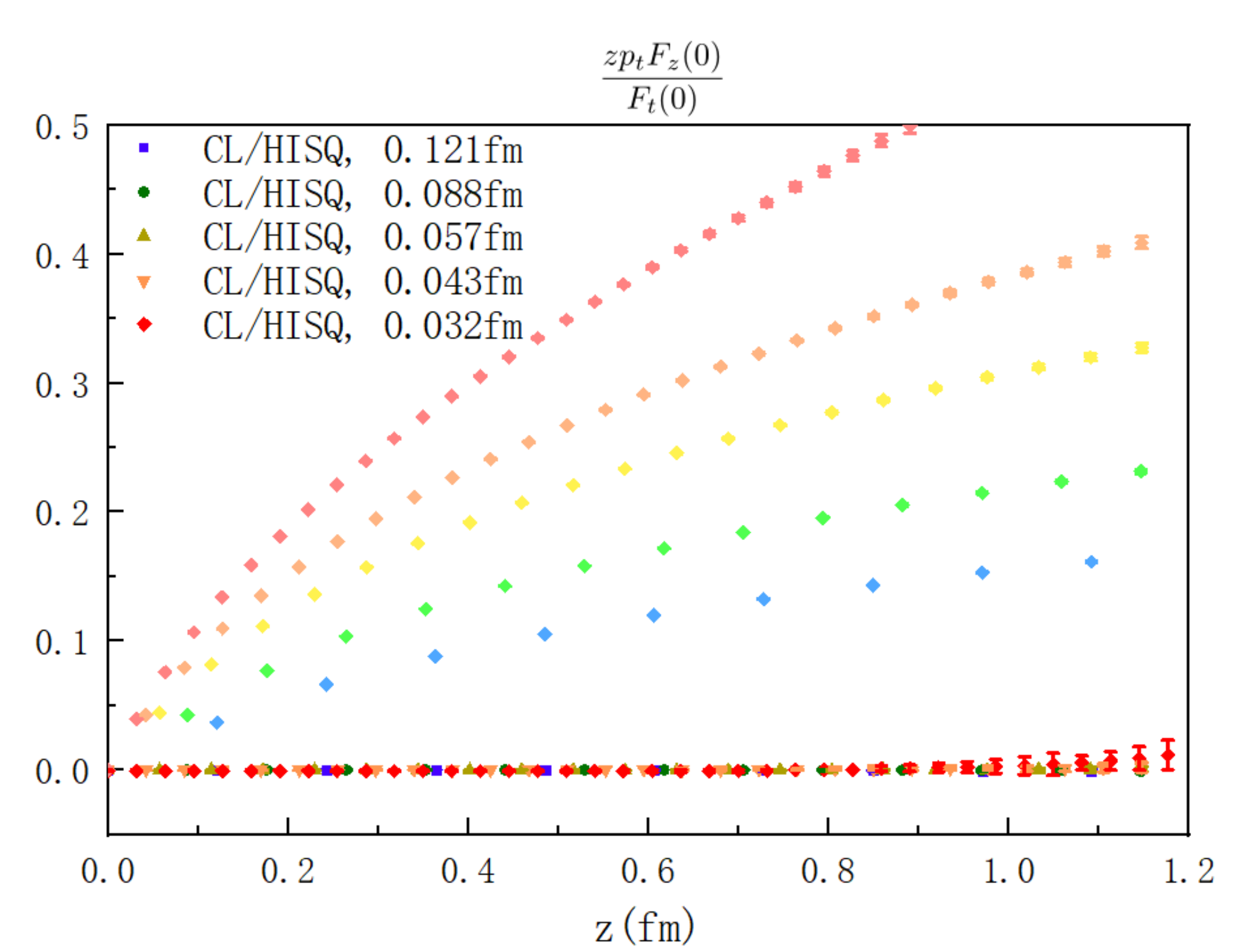}
   \includegraphics[width=8cm]{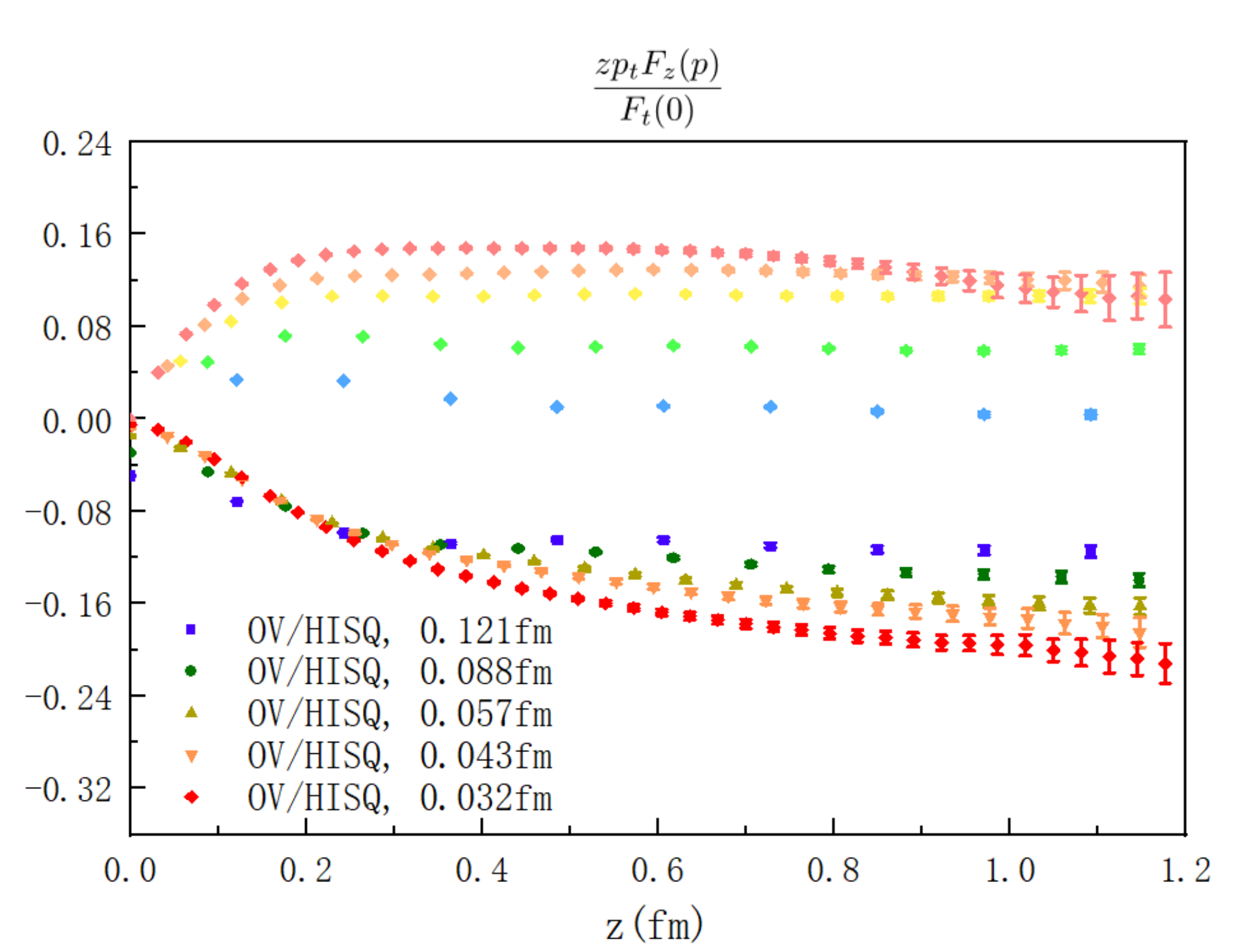}
   \includegraphics[width=8cm]{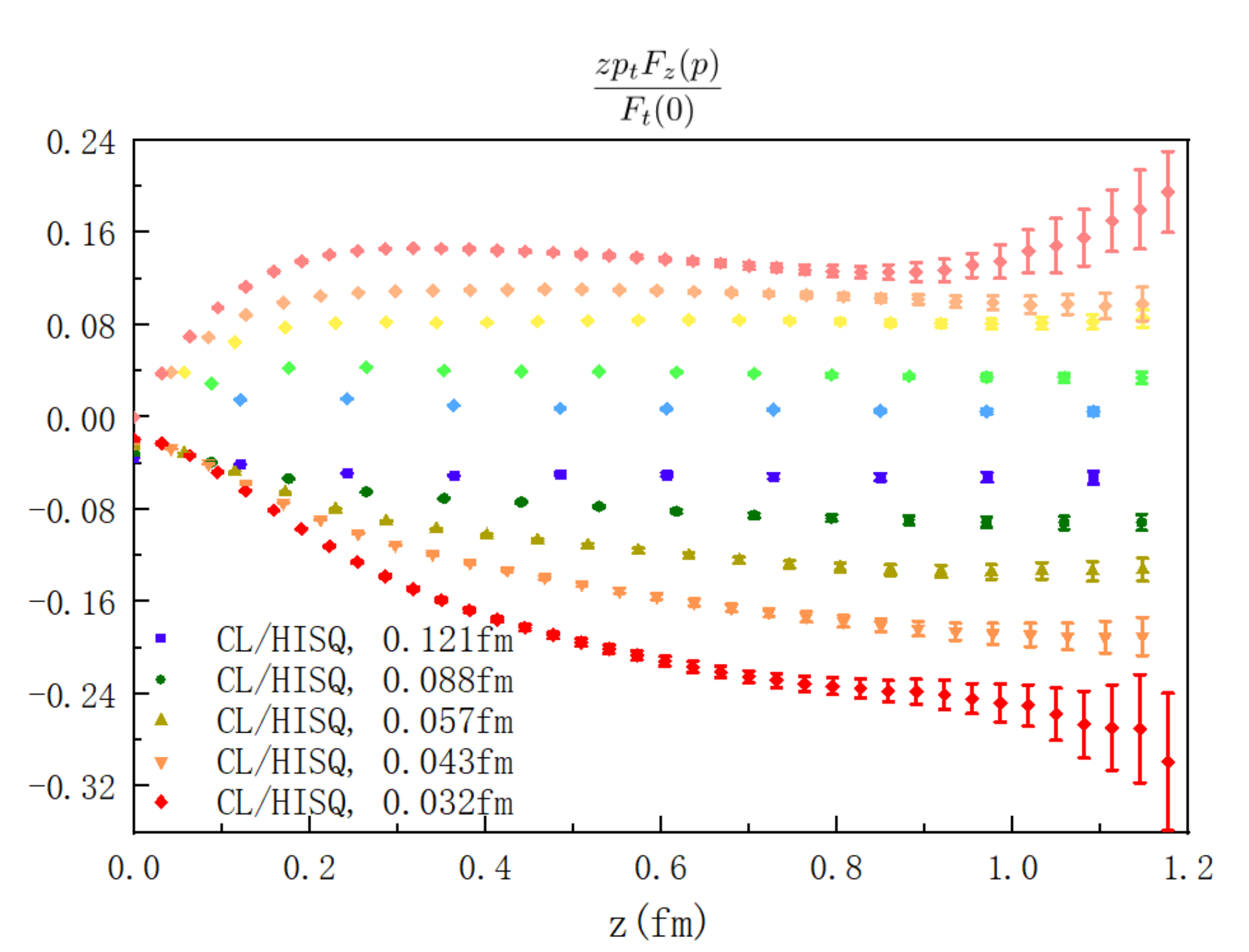}
  \caption{Similar to Fig.~\ref{fig:mixing_terms2} but for the $\tilde{F}_{p}(0;z)$ case.}
  \label{fig:mixing_terms3}
\end{figure}

The last part is the $zp_t\tilde{F}_{z}(0;z)$ term which can be understood understood using similar arguments. As shown in the upper panels of Fig.~\ref{fig:mixing_terms} for the $p_z=0$ case, the values are about 30\% of $F_t$ and then the linear divergence dominates  the lattice spacing dependence; but for  $p_z\neq 0$, shown in the lower panel, the values are much smaller ($\sim$ 10\% of $F_t$) and then both linear divergence and discretization error contribute equally to the lattice spacing dependence.

Thus, using the RI/MOM renormalization constant defined in Eq.~(\ref{eq:z_pslash}) with the p-slash projection, will introduce an additional lattice spacing dependence due to the linear divergence and discretization errors, which can result in a complicated continuum extrapolation. 

\begin{figure}[tbph]
  \centering
   \includegraphics[width=8cm]{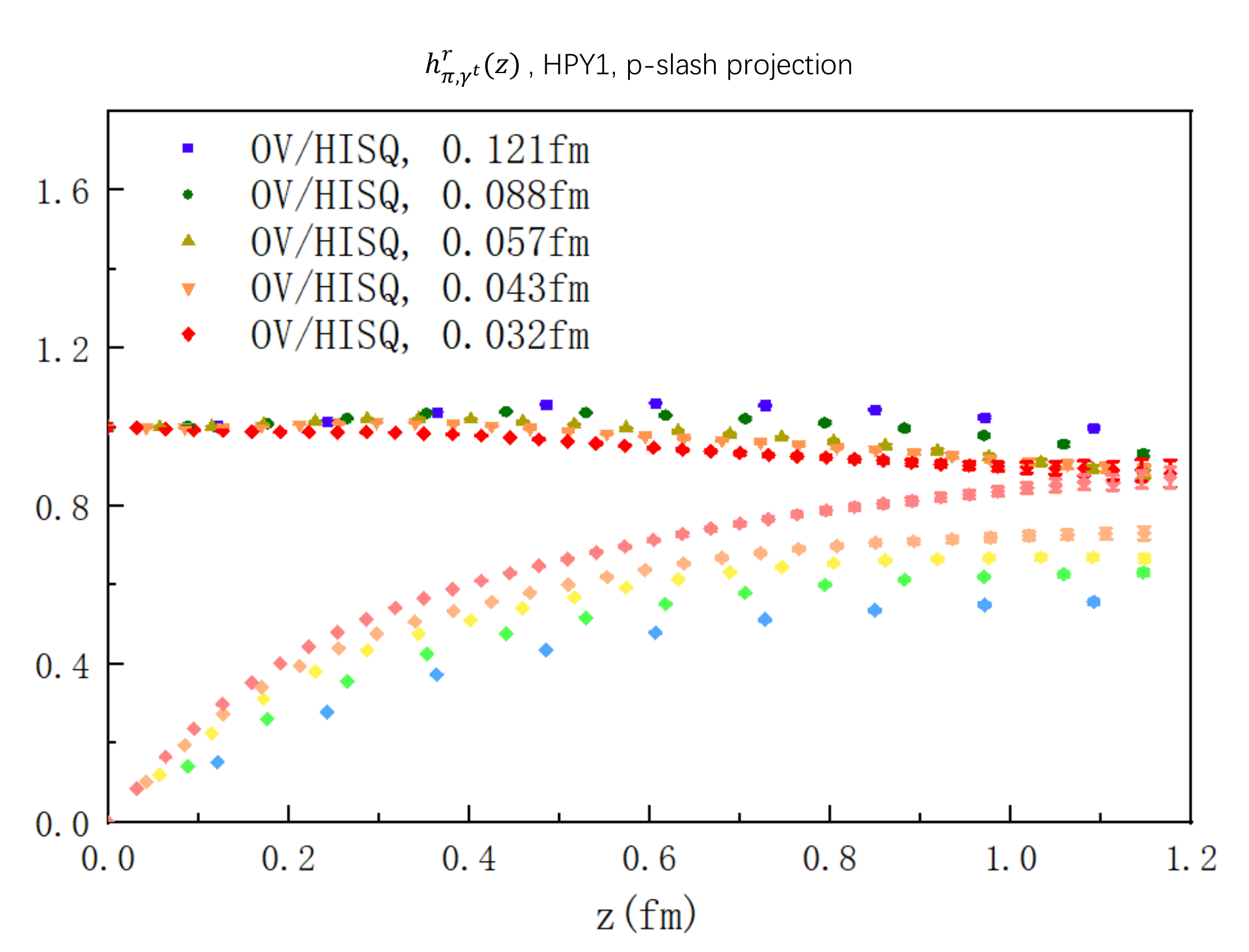}
   \includegraphics[width=8cm]{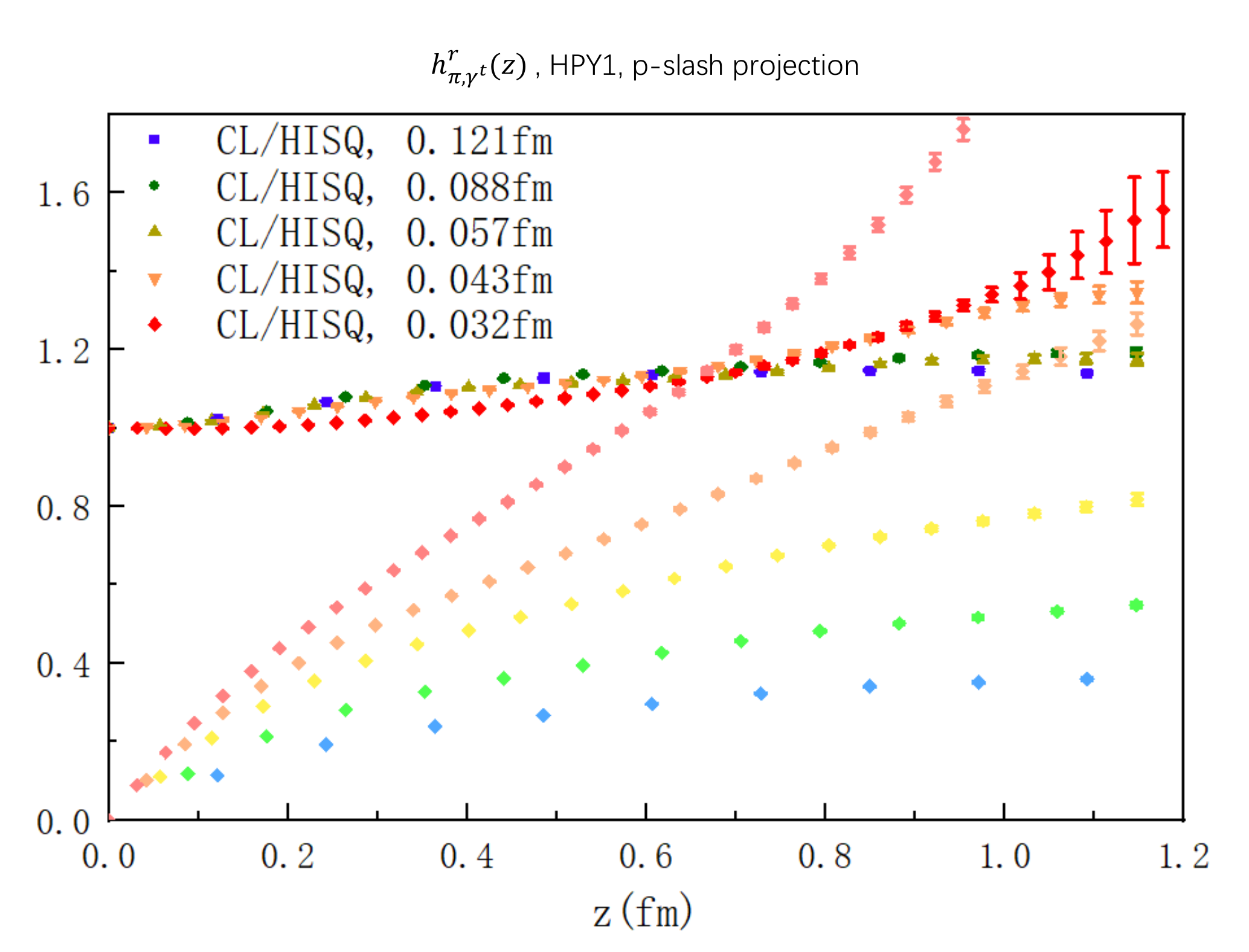}
  \includegraphics[width=8cm]{figures/overlap_milc_hyp1.pdf}
  \includegraphics[width=8cm]{figures/milc-clover-old-new-pion-hyp1.pdf}
  \caption{The RI/MOM renormalized pion matrix elements $h^r_{\pi,\gamma^t}(z)$ using the overlap (OV, upper left panel) or clover (CL, upper right panel) fermion on HISQ sea ensembles. The cases using the minimum projection are also shown in two lower panels for comparison. The darker data points correspond to the real parts and lighter ones are the imaginary parts. It is funny that the lattice spacing dependence of $h^r_{\pi,\gamma^t}(z)$ with the p-slash projection in the clover fermion case is majorly moved to the imaginary part.}
  \label{fig:p-slash}
  \end{figure}

Eventually, in Fig.~\ref{fig:p-slash}, we show the RI/MOM renormalized pion matrix elements $h^r_{\pi,\gamma^t}(z)$ with the p-slash projection and $p=2\pi(0,0,5,5)/L$, for both overlap fermions (upper left panel) or clover fermions (upper right panel) on HISQ sea ensembles. The cases with the minimum projection (the $\gamma_t$ projector with $p_z=p_t=0$) are also presented in the two lower panels for comparison. As in the left two panels, the difference between $h^r_{\pi,\gamma^t}(z)$ at the smallest and largest lattice spacing with the p-slash projection is much larger than that for the minimum projection and overlap fermions. Thus, it is not surprising that the clover case can behave even worse, as is illustrated in the right two panels. While the real part of $h^r_{\pi,\gamma^t}(z)$ using the p-slash projection is somehow similar for different lattice spacings when $z<0.8$ fm, the imaginary part shows an obvious linear divergence remnant which cannot be removed by perturbative matching.

\end{widetext}

\end{document}